\documentclass[lettersize,journal]{IEEEtran}

\usepackage{float}
\usepackage{graphicx}
\usepackage{clrscode}
\usepackage{stfloats}
\usepackage{paralist}
\usepackage{tikz}
\usepackage{soul} 
\usepackage{color}
\usepackage{cite}
\usepackage{comment}
\usepackage{amsmath, amssymb, mathrsfs, amsfonts}
\usepackage{amsthm}
\usepackage{mathtools}
\usepackage{epsfig, epstopdf}
\usepackage{subfig}
\usepackage{algorithm}
\usepackage{enumitem}
\usepackage{algorithmic}
\usepackage{hyperref}
\usepackage{lipsum}
\usepackage{pbox}
\usepackage{array}
\usepackage{tablefootnote}
\usepackage[square,numbers,sort&compress]{natbib}
\usepackage[font={footnotesize}]{caption}

\usetikzlibrary{arrows,automata}

\allowdisplaybreaks

\newtheorem{theorem}{Theorem}
\newtheorem{lemma}{Lemma}

\newtheorem{corollary}{Corollary}

\newtheorem{remark}{Remark}

\newcommand{\E}{\mathbb{E}}

\newcommand{\argmax}{\arg\!\max}

\newcommand{\RNum}[1]{\uppercase\expandafter{\romannumeral #1\relax}}

\begin{document}
\nocite{*}

\title{Physical Layer Security in Random NOMA-Enabled Heterogeneous Networks}
\author{Elmira Shahraki, Mahtab Mirmohseni, and Hengameh Keshavarz
\thanks{E. Shahraki and H. Keshavarz are with the Department of Electrical  and Computer Engineering, University of Sistan and Baluchestan, Zahedan, Iran (email: {elmira.shahraki}@pgs.usb.ac.ir; {keshavarz}@ece.usb.ac.ir).

M. Mirmohseni is with the Institute for Communication Systems (ICS) Department of Electrical Engineering, University of Surrey, Guildford, United Kingdom (email: m.mirmohseni@surrey.ac.uk).}}

\maketitle

\vspace{-1.5cm}

\begin{abstract}
The performance of physical layer secrecy approach in non-orthogonal multiple access (NOMA)-enabled heterogeneous networks (HetNets) is analyzed in this paper. A $K$-tier multi-cell HetNet is considered, comprising NOMA adopted in all tiers. The base stations, legitimate users (in a two-user NOMA setup), and passive eavesdroppers,  all with single-antenna, are randomly distributed. Assuming independent Poisson point processes for node distribution, stochastic geometry approaches are exploited to characterize the ergodic secrecy rate. A lower bound on the ergodic secrecy rate along with closed-form expressions for  the lower bound on the ergodic rates of the legitimate users in a special case are derived. Moreover, simpler expressions for the ergodic secrecy rate are obtained in the interference-limited regime with vanishing noise variance. The effect of multi-tier technology, NOMA, and physical layer secrecy are investigated using numerical results. The results reveal that applying HetNet to a secure multi-cell NOMA system improves the spectrum efficiency performance.  
\end{abstract}


\section{Introduction}
Nowadays, cellular systems focus on beyond 5G networks, where to meet high spectral efficiency, important new technologies such as heterogeneous networks (HetNets) and non-orthogonal multiple access (NOMA) are proposed \cite{5GN2015}-\cite{5GMAN2016}. HetNets are typically composed of macrocells with high transmission power and coverage. Macrocell, in turn, overlays smallcells such as picocells and femtocells which have low transmission power and coverage. Thus, network capacity is improved and the coverage is extended due to distance reduction between end users and access nodes. Channel reuse is also increased by deploying small base stations (BSs). Consequently, traffic can be offloaded to smallcells to target a higher spectral efficiency using HetNets \cite{HetNetANJV2010}, \cite{HetNetAJYTTMTOD2011}. On the other hand, the basic idea of NOMA is to allocate the same time and frequency resources to different users, instead of using orthogonal spectrum. Therefore, significant improvements in spectral efficiency can be achieved by NOMA. Particularly, power-domain NOMA improves spectral efficiency by superposing multiple users in the power domain. This can be obtained by superposition coding (SC) at the transmitter and successive interference cancellation (SIC) at the receivers \cite{NOMAZXGRJV2017}, \cite{NOMALBZZSL2018}.

Due to the broadcasting nature of transmission medium, security is an essential feature in wireless communication networks. The information theoretic methods to provide secrecy at the physical layer become popular in beyond 5G networks. Thanks to their easier key management and resistance against computationally powerful eavesdroppers \cite{PLSNLGMJM2015}, \cite{PLSYACGKX2018}. Physical layer security (PLS) methods exploit the physical characteristics of wireless channels including noise, fading, and interference  \cite{PLSYHL2016}, \cite{PLSTXH2017}. In this paper, we study the performance of two beyond 5G key technologies, HetNet and NOMA, in the presence of PLS approaches.

The existing literatures on secure NOMA system \cite{NOMASECYHQZ2016}-\cite{NOMASECBMCXFJF2018}, secure HetNet  \cite{HetnetSECHTJDM2016}-\cite{HetnetSECY2018}, and NOMA-based HetNet \cite{HetnetNOMAYZMAJ2017}-\cite{HetnetNOMATJXSQZK2018} either optimize the system parameters for a single shot of a network or evaluate the performances of HetNet, NOMA, and  PLS, separately, in random networks (i.e., a network with randomly located nodes). However, the PLS in NOMA-enabled HetNet in random networks had not been considered, which is the main focus of this paper. Note that providing PLS in HetNets and also multi-cell NOMA systems result in spectrum efficiency reduction. Hence, the motivation of studing  NOMA-enabled HetNet under secrecy constraint is to measure this reduction. 
\subsection{Related Works}
PLS in NOMA-based systems was studied from various perspectives \cite{NOMASECYHQZ2016}-\cite{NOMASECBMCXFJF2018}. Optimization approaches were taken in \cite{NOMASECYHQZ2016} and \cite{NOMASECBANV2017}. 
% maximizing the secrecy sum rate rate in a single-input single-output NOMA system by satisfying users' quality of service (QOS) was investigated. The system includes one transmitter, several legitimate users, and one eavesdropper. A closed-form expression for the optimal power allocation policy is obtained.
 The security in NOMA large scale networks was studied in  \cite{NOMASECYZMYL2017} using stochastic geometry to calculate the secrecy outage probability, where the network consists of one BS, several legitimate users, and  eavesdroppers in both single-antenna and multiple-antenna scenarios. %In the single-antenna scenario, a user area around the BS and a protected zone around the BS as an exclusion area of eavesdropper is assumed. In the multiple-antenna scenario, artificial noise (AN) is generated at the BS to further improve security and reduce SOP. 
 %Minimizing transmission power in a NOMA system involves one transmitter, several legitimate users and one eavesdropper was studied in \cite{NOMASECBANV2017}. QOS per user, which is the real rate received per user, and SOP constraints  are considered. Also, maximizing the minimum confidential information rate among users with transmission power and SOP constraints is  examined.
 %The study of security in a single-cell multiuser downlink communication system was discussed in \cite{NOMASECXZCDR2018}. In this structure, the BS is equipped with massive multiple-input multiple-output (MIMO) technology, and users are grouped into clusters that there is one eavesdropper in each cluster. NOMA and non-orthogonal channel estimation techniques are used to improve the signal quality of users as well as inter user interference to confuse eavesdroppers and thus improve the secrecy performance. The closed-form is computed for the ergodic secrecy rate and there are two optimization problems to maximize the secrecy rate and minimize the transmission power.
 %@article{NOMASECXZCDR2018,
 %title={{Exploiting} {Inter-User} {Interference} for {Secure} {massive} {Non-Orthogonal} {Multiple} {Access}},
 %author={Chen, Xiaoming and Zhang, Zhaoyang and Zhong, Caijun and Ng, Derrick Wing Kwan and Jia, Rundong},
 %journal={IEEE Journal on Selected Areas in Communications},
 %volume={36},
% number={4},
 %pages={788--801},
% year={2018},
 %publisher={IEEE}
%} 
 Secrecy outage probability and strictly positive secrecy rate were analyzed in \cite{NOMASECJLM2018}, which investigate of a cooperative NOMA system with a single relay, one BS, and an eavesdropper. %Also, SOP is studied in high  signal-to-noise ratio (SNR) and it is shown that  the secrecy function of AF and DF in high SNR is the same.
  The PLS analysis in a two-way channel with a trusted multiple-antenna relay in the presence of  eavesdroppers was studied in \cite{NOMASECBMCXFJF2018}. Artificial noise and full-duplex techniques were used at the relay to improve the secrecy performance.  A closed-form expression for the ergodic secrecy rate was obtained in both single eavesdropper and multiple eavesdroppers cases. %The communication consists of a multiple-access phase and a broadcast phase.  
The ergodic secrecy rate analysis in the \textit{multi-cell} NOMA systems has not been taken into account, which is addressed in this paper.

 Exploiting PLS in HetNets for random networks was studied in \cite{HetnetSECHTJDM2016}-\cite{HetnetSECY2018}. In \cite{HetnetSECHTJDM2016}, secrecy and connection probabilities along with sum secrecy  rate were studied in a multi-tier heterogeneous cellular network. The position of BSs, legitimate users, and eavesdroppers are characterized by homogeneous Poisson point processes (HPPPs). Each BS employs the artificial noise transmission strategy and the user association policy is based on  the truncated average received signal power. %Secrecy probability and connection probability are studied, and  network-wide secrecy throughput and minimum per user secrecy throughput are studied  subject to Secrecy probability and connection probability.
 A dynamic coordinated  multi-point transmission scheme is introduced in \cite{HetnetSECMXFH2016} for BS selection in heterogeneous cellular networks, where the received signal power for legitimate users are used in BS selection process. The secure coverage probability was calculated by considering co-channel interference and worst-case scenario for eavesdroppers. The area ergodic secrecy rate, the secrecy outage probability, and the energy efficiency in heterogeneous cloud radio access network (RAN) were studied in \cite{HetnetSECLKMAS2016} considering soft fractional frequency reuse (S-FFR), where two-tier heterogeneous cloud RAN consists of massive MIMO macrocell BSs in the first tier and remote radio heads in the second tier. Locations of the macro BSs, the remote radio heads, and the passive eavesdroppers were modeled as HPPPs. In \cite{HetnetSECWKK2017}, artificial noise-aided PLS in multi-antenna smallcell networks was investigated. %In high-load cell mode, a multi-user zero-forcing beamforming transmission scheme is introduced to improve sum secrecy throughput.
   Closed-form expressions for the connection and the secrecy outage probabilities were obtained as well as a semi closed-form  lower bound on the average secrecy rate. In \cite{HetnetSECSYCNG2018}, a user association based on the maximum secrecy capacity in two-tier heterogeneous cellular networks with in-band interference was studied. %Considering the in-band interference, a user connects to the BS that provides the largest average secrecy capacity  under the max-SC association.
    Keeping the user association scheme in mind, connection and secrecy probabilities and network secrecy throughput were analyzed. PLS was also studied for a heterogeneous spectrum-sharing cellular network in \cite{HetnetSECY2018}. %considering overlay spectrum sharing, interference-limited underlay spectrum sharing, and interference-canceled USS methods . 
    The network includes a macro BS  and a small BS  that send messages to legitimate macro and small users in the presence of an eavesdropper. %Both methods overlay spectrum sharing (OSS) and interference-limited underlay spectrum sharing (USS) are studied for access to the same spectrum and  interference-canceled USS (IC-USS) method is proposed to improve transmission security. %In IC-USS, which transmits macro BS and small BS simultaneously over the same transmission spectrum, a specific signal is designed to reduce mutual interference between MU and SU and this signal is extremely  harmful to  the eavesdropper.
     Overall outage  and intercept probabilities were obtained in closed-form and secrecy diversity analysis was performed to evaluate performance. %Secrecy and coverage performance were analyzed by considering pilot attacks from active eavesdroppers and pilot contaminations in a hybrid sub-6 GHz massive MIMO and millimeter wave two-tier HetNet \cite{HetnetSECWKSK2018}. The mmWave tier performed better in terms of secrecy and coverage performance than the sub-6 GHz tier through densifying the BSs.

Without secrecy constraint, the outage probability and the ergodic rate of the HetNet-NOMA systems in random networks were studied  in \cite{HetnetNOMAYZMAJ2017}-\cite{HetnetNOMATJXSQZK2018}.
 %@article{HetnetNOMAMPNMA2018,
 %title={{Optimal} and {Fair} {Energy} {Efficient} {Resource} {Allocation} for {Energy} {Harvesting-Enabled-PD-NOMA-Based} {HetNets}},
 %author={Moltafet, Mohammad and Azmi, Paeiz and Mokari, Nader and Javan, Mohammad Reza and Mokdad, Ali},
 %journal={IEEE Transactions on Wireless Communications},
 %volume={17},
 %number={3},
 %pages={2054--2067},
 %year={2018},
 %publisher={IEEE}
 %}
 %The optimization approache is tacken in \cite{HetnetNOMAMPNMA2018}.  In a NOMA-based HetNet, different fairness methods such as max–min, proportional and minimum delay potential are studied \cite{HetnetNOMAMPNMA2018}. Two SIC ordering methods based on Channel to Noise Ratio (CNR) and Channel to interference gain plus noise ratio (CINR) are suggested. Problem formulations are proposed based on the above fairness methods  to maximize energy efficiency by considering energy harvesting. An iterative algorithm is considered to solve the proposed problem, which the main problems are decomposed into two  subcarrier assignment problem and power allocation problem at each iteration.
 In \cite{HetnetNOMAYZMAJ2017}, the coverage probability, the ergodic rate, and the energy efficiency in two-tier HetNets were analyzed. The first and the second tiers were equipped with massive MIMO and NOMA technologies, respectively. %The user association is based on the maximum average received power in each tier.The  analytical  expression for the coverage probability is obtained in the second tier. Exact analytical expressions for ergodic rate in the second tier and a tractable lower bound  for ergodic rate in the first tier are obtained.
 %It can be seen that increasing the number of antennas in the first tier significantly improved ergodic rate and had an opposite effect on energy efficiency. The second tier achieved higher energy efficiency than the first tier. 
 The coverage probability and achievable rate were analyzed in a downlink NOMA-based HetNet in \cite{HetnetNOMACD2018}. %The non-coordinated NOMA method, in which no void BSs are coordinated to help a non-void BS jointly transmit its NOMA signals, is analyzed. 
 For improvement of NOMA and SIC, a coordinated joint transmission NOMA method was introduced. %The optimal allocation of transmit powers to NOMA users for maximize cell coverage and cell throughput is also analyzed.
 %The analysis of outage probability, offloading and NOMA compatibility probabilities in a two-tier HetNet were studied in \cite{HetnetNOMAPVST2018}. The second tier was equipped with carrier sensing for interference management (using repulsive point processes model) and NOMA. %The second tier supports offloaded users from congested first tier for load balancing, and the offloaded first tier user  pairs  with a available user for NOMA. The analysis of outage probability for the first and second tiers, offloading and NOMA compatibility probability is performed.
 Analysis of the coverage probability  and the spectral efficiency  in NOMA-based HetNets were studied in \cite{HetnetNOMACM2018} regarding interference coordination. Two well-known methods, namely strict fractional frequency reuse and soft frequency reuse, were used to reduce inter-cell interference. 
 % for both open-access and closed-access options of HetNets is also analyzed.
 The coverage probability  and the achievable rate were characterized in \cite{HetnetNOMATJXSQZK2018} for a NOMA-based two-tier HetNet with non-uniform smallcell deployment. %For the purpose of non-uniform deployment and to establish coverage and energy efficiency, small BSs within a certain distance from the macro BSs are not active and go to a asleep. The proposed NOMA user pairing  resulted in higher achievable rate than the random user pairing.

 Optimization approaches for secure NOMA-based HetNets were also studied in \cite{HetnetNOMASECYXLXZD2019} and \cite{HetnetNOMASECMPNKH2019}. Cooperative jamming was utilized in a two-tier HetNet in \cite{HetnetNOMASECYXLXZD2019}. The first tier and the second tier were equipped with massive MIMO and NOMA technologies, respectively. The proposed algorithms are presented to maximize the secrecy rate of target users subject to the QoS constraints of other users. %Two assumptions are considered to be imperfectly known of CSI and collusion of multiple eavesdroppers. %To overcome the challenges created by these assumptions, the three secrecy transmission algorithms, robust beamforming algorithm, robust power allocation algorithm, and robust joint optimization algorithm, have been proposed to be respectively applied in the first tier, the second tier, and both tiers of HetNet. 
In \cite{HetnetNOMASECMPNKH2019}, a resource allocation algorithm (joint subcarrier and power allocation) was studied in a NOMA based two-tier HetNet. The network consisted of one macro BS and multiple small BSs in both single-antenna and multiple-antennas modes. %A resource allocation algorithm (joint subcarrier and power allocation) is proposed to maximize the sum secrecy rate in which the eavesdroppers are not allowed to perform SIC, but the legitimate users are able to perform it. Two modes,  perfect and imperfect CSI availability of the legitimate users and the eavesdroppers at BSs, are analyzed. The proposed technique is evaluated in heterogeneous ultra dense networks for secure massive connectivity.
Unlike \cite{HetnetNOMASECYXLXZD2019} and \cite{HetnetNOMASECMPNKH2019}, we study the performance of secure NOMA-based HetNet in random networks.
\subsection{Contributions}
To the best of our knowledge, there is no study on PLS in random NOMA-enabled HetNets. Moreover, the ergodic secrecy rate in \textit{multi-cell} NOMA systems has not been investigated. In addition to the improved spectral efficiency, both NOMA and HetNet may contribute in realizing PLS technique. The interference caused by macrocells and smallcells, one of the main challenges in HetNets, corrupts the eavesdroppers signals in a friendly manner. Furthermore, NOMA causes interference in the network which is beneficial to provide system security. However, the challenges of analyzing NOMA-based HetNet in random networks with PLS approach fall into the complexity of calculations and the difficulty of deriving closed-form secrecy rate expressions. Our main contributions are as follows:

\tikz\draw[black,fill=black] (0,0) circle (.5ex);  The secrecy performance of a $K$-tier multi-cell HetNet with  NOMA enabled in all tiers is investigated, exploiting stochastic geometry approaches. The studied $K$-tier multi-cell HetNet  consists of single-antenna  BSs in all tiers; the legitimate users employ a two-user NOMA approach, and the eavesdroppers passively intercept the secure messages. The locations of BSs, legitimate users, and eavesdroppers are  randomly distributed regarding HPPP. The model is described in Section II. 
	
\tikz\draw[black,fill=black] (0,0) circle (.5ex); Analytical expressions for the ergodic secrecy rate of the secure NOMA-based HetNet (SN-Het) is derived in Section III. To this end, user association probability, probability density function (PDF) of users' distances, SINR analysis considering NOMA transmission, and the characteristic function of interference are characterized. Although the closed-form of ergodic secrecy rate is not reached for the general case, the results are easily computable. We also derive lower bounds on the ergodic rates of the legitimate users. The lower bounds are reduced to closed-form in a special case with the same per-tier path loss exponents and no biasing. In addition, simpler expressions for the ergodic secrecy rate in  the interference-limited regime is achieved considering the same path loss exponents, zero noise power, and no biasing for all tiers.
	
\tikz\draw[black,fill=black] (0,0) circle (.5ex); Numerical and simulation results are  provided in Section IV to evaluate the performance of considered SN-Het. The derived ergodic secrecy rate is compared with rates in HetNet \cite{HetnetHYPJ2012}, HetNet-NOMA \cite{HetnetNOMAYZMAJ2017}, secure HetNet \cite{HetnetSECLKMAS2016}, and secure multi-cell NOMA (assuming only one tier in the derived results of ssection III for SN-Het) to observe the effects of secrecy constraint as well as, NOMA,  and HetNet technologies. The results show that applying HetNet to a secure multi-cell NOMA system improves the spectrum efficiency, while the secrecy constraint degrades the ergodic rate as expected. It is also inffered that increase in number of network tiers with a fixed density of BSs results in the ergodic secrecy rate improvement. On the other hand in a two-tier network, the ergodic secrecy rate is improved by increasing the density of second tier BSs that results in proper interference for security. 

\section{Problem Description}
\begin{figure}
	\centering
	\vspace{-0.0cm}
	  	\includegraphics[width=8cm,height=8cm]{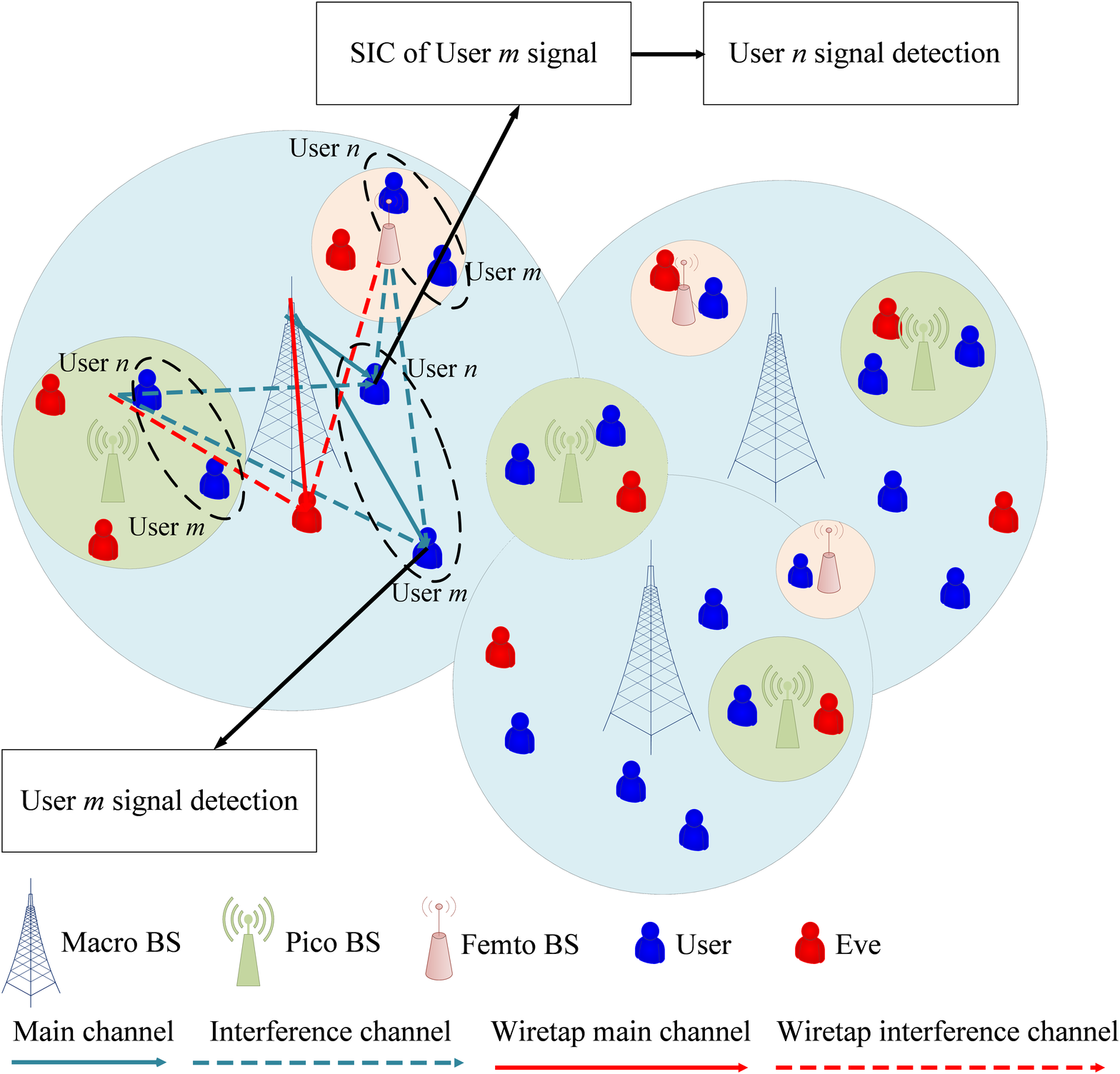}  
	\vspace{-0.1cm}
	\caption[The SN-Het system model]{Illustration of the SN-Het system model\footnotemark.}
	\label{Fig.fig1}
	\vspace{-0.5cm}
\end{figure}

  As shown in Fig. \ref{Fig.fig1}, a $K$-tier multi-cell HetNet with macrocells in the first tier and smallcells in remaining tiers is considered. Users are randomly distributed according to an HPPP $\Phi_u$ with intensity $\lambda_u$ and we exploit two-user NOMA technique. A number of passive eavesdroppers are randomly distributed throughout the network to intercept the secrecy messages. The positions of eavesdroppers and BSs in the $j$-th tier are modeled according to  HPPPs denoted as 
 $\Phi_e$
 and
 $\{\Phi_{b_{j}}\}_{j=1,...,K}$
  with intensities
  $\lambda_e$
  and
$\lambda_j$, 
respectively.  $\Phi_{e_k}$ is a set of eavesdroppers in the  tier $k$. Users, eavesdroppers, and BSs are equipped with a single-antenna. It is assumed that the channel state information (CSI) of users at the BS is known. However, the CSI of eavesdropper is unknown at the BS, and only the knowledge of its channel distribution  is available. Accordingly, the ergodic secrecy rate is computed in this paper.
\footnotetext{For brevity, the interference of BSs in macrocells and the intersection of picocells and femtocells are not shown, though they are considered in the mathematical model.}

\subsection{Signal Model}\label{IA}
Each BS communicates with a set of users in the presence of eavesdroppers. The set of all users at $q$-th BS in the $k$-th tier, ${\text {BS}}_{k,q}$, is defined as $\mathcal{N}_{k,q}=\{1,2,...,{{n}_{k,q}}\}$ where ${{n}_{k,q}}$ is a random variable. Without loss of generality, ${\text {BS}}_{k,q}$ divides $\mathcal{N}_{k,q}$ into $\frac{{{n}_{k,q}}}{2}$ subsets, to enable a two-user NOMA scheme with random pairing in each subset. We consider one of the subsets, including users $m$ and $n$. It is also assumed that user $m$ is farther from ${\text {BS}}_{k,q}$ than user $n$. 
Employing NOMA scheme, ${\text {BS}}_{k,q}$ transmits signal 
$\sqrt{{a_{m_{k,q}}}P_k}s_{m_{k,q}}+\sqrt{{a_{n_{k,q}}}P_k}s_{n_{k,q}}$, where 
$s_{m_{k,q}}$
and 
$s_{n_{k,q}}$
are the transmitted messages for user $m$ and user $n$, respectively. 
$P_k$
is the transmit power of ${\text {BS}}_{k,q}$;
$(a_{m_{k,q}},a_{n_{k,q}})$ are the power allocation coefficients for users $m$ and $n$, while ${a_{m_{k,q}}}\ge{a_{n_{k,q}}}$ and  ${a_{m_{k,q}}}+{a_{n_{k,q}}}=1$. The received signals at user $m$ and user $n$ and also the $e$-th  eavesdropper are as:
\begin{align}\label{eq:01}
&{y_{i_{k,q}}}={h_{i_{k,q}}}\left(\sqrt{{a_{m_{k,q}}}P_k}s_{m_{k,q}}+\sqrt{{a_{n_{k,q}}}P_k}s_{n_{k,q}}\right)+{z_{i_{k,q}}}\nonumber\\ 
&+\sum_{j=1}^K\hspace{0.03cm}\sum_{l\in\Phi_{b_j}\backslash\left\{BS_{k,q}\right\}}\hspace{-0.22cm}{h_{i_{j,l}}}\left(\sqrt{{a_{m_{j,l}}}P_j}s_{m_{j,l}}+\sqrt{{a_{n_{j,l}}}P_j}s_{n_{j,l}}\right),
\end{align} 
 where  $i\in\{m,n,e\}$;  
 ${h_{i_{k,q}}}=\sqrt{{g_{i_{k,q}}}}{d_{i_{k,q}}}^{-{{\alpha_k}/2}}$
 is the channel coefficient between ${\text {BS}}_{k,q}$ and the $i$-th user/eavesdropper in $\mathcal{N}_{k,q}$/$\Phi_{e_k}$; 
 ${g_{i_{k,q}}}\sim \exp(1)$
 denotes the small-scale fading transmission/eavesdropping channel power gain,
 ${d_{i_{k,q}}}$
 stands for the distance between $i$-th user/eavesdropper and ${\text {BS}}_{k,q}$, and $\alpha_k$
 is the path loss exponent of the $k$-th tier.
  Moreover,
  ${h_{i_{j,l}}}=\sqrt{{g_{i_{j,l}}}}{d_{i_{j,l}}}^{-{{\alpha_j}/2}}$
   is the channel coefficient between  ${\text {BS}}_{j,l}$ ($l$-th BS in tier $j$) and $i$-th  user/eavesdropper in $\mathcal{N}_{k,q}$/$\Phi_{e_k}$ where
   ${g_{i_{j,l}}}\sim \exp(1)$
   is the small-scale fading channel gain; 
   ${d_{i_{j,l}}}$
   denotes the distance between $i$-th user/eavesdropper  and ${\text {BS}}_{j,l}$. ${z_{i_{k,q}}}$
stands for the complex additive white Gaussian noise (AWGN) with zero-mean and variance
$\sigma^2$,
${z_{i_{k,q}}}\sim{\mathcal{CN}}(0,\sigma^2)$.

 \subsection{User Association Statistics and Distance Distribution}\label{IA}
 The maximum biased average received power of each tier is assumed for user association \cite{HetnetHYPJ2012}. Keeping the $w$-th user, at tier $k$, as the associated user in mind, average biased received power is:
 \begin{align}\label{eq:02} 
 P_{r_{k,q}}=a_{w_{k,q}}P_kB_k {R_u}_{k,q} ^{-\alpha_k},
 \end{align}
 where $w\in\{m,n\}$, 
 %, $u\in\{f,s\}%
 $B_k$
 is a positive bias factor of tier $k$ (the bias factor of all BSs in tier $k$ is the same).
 ${R_u}_{k,q}$
 denotes the distance between a typical user and ${\text {BS}}_{k,q}$.
 
 %\noindent
According to \cite[Lemma~1]{HetnetHYPJ2012}, the probability of a typical user association with the secure NOMA-based HetNet BSs in tier $k$ is:
\begin{align}\label{eq:03}
{A_{k}}={2\pi\lambda_{k}}\int_{0}^{\infty} {r}\exp\Big\{-\pi\sum_{j=1}^K{\lambda_{j}(\hat{P_j}\hat{B_j})^{2/\alpha_j}}{{r}^{2/\hat\alpha_j}}\Big\}\,dr, 
\end{align} 
where 
$\hat{P_j}=\frac{P_j}{P_k}$,
$\hat{B_j}=\frac{B_j}{B_k}$,
and
$\hat{\alpha_j}=\frac{\alpha_j}{\alpha_k}$
are the ratios of the interfering $j$-th tier to the serving $k$-th tier.     
%, where $u\in\{f,s\}$
Regarding \cite[Lemma~3]{HetnetHYPJ2012}, the PDF of distance $R_{u} = \argmax\limits_{k \in \{1,2,...,K\}}(P_{r_{k,q}})$ between a typical user and its serving BS in the $k$-th tier is as follows:
\begin{align}\label{eq:04}
{f_{R_u}(r)}=\frac{2\pi\lambda_{k}}{A_{k}} {r}\exp\Big\{-\pi\sum_{j=1}^K{\lambda_{j}(\hat{P_j}\hat{B_j})^{2/\alpha_j}}{{r}^{2/\hat\alpha_j}}\Big\}. 
\end{align} 
	
	%{\arg\!\max_{f\in\Phi_{b_j}}}{P_{r_{f}}}
\subsection{NOMA Transmission and SINR Analysis}\label{IA}
\begin{figure}
\centering
\vspace{-0.4cm}
%	\subfloat[Case\RNum{1}]{\includegraphics[width=7cm]{A.png}}
	\subfloat[Case \RNum{1}]{\includegraphics[width=4cm,height=4cm]{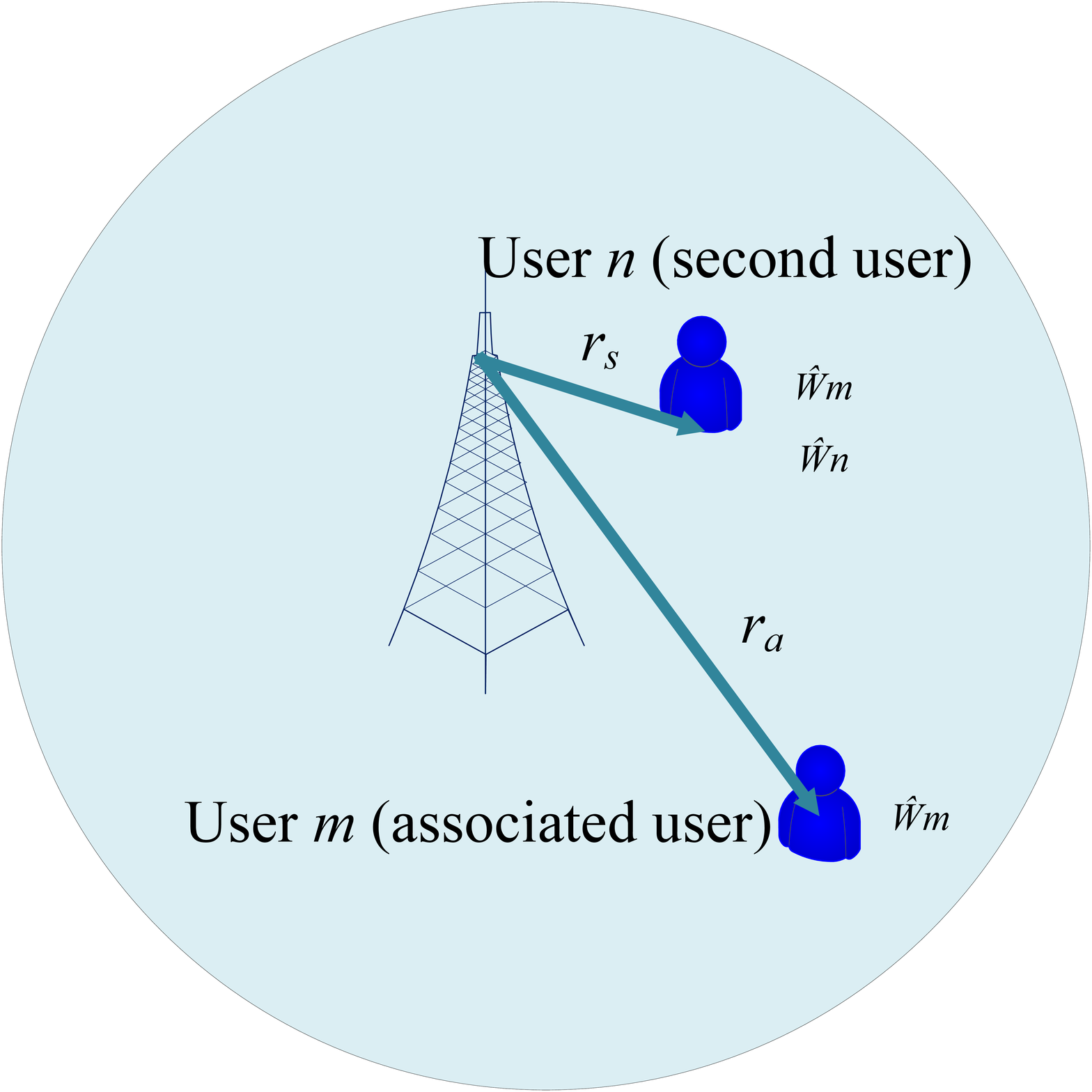}}
	\qquad
    %\subfloat[Case\RNum{2}]{\includegraphics[width=7cm]{B.png}}
	\subfloat[Case \RNum{2}]{\includegraphics[width=4cm,height=4cm]{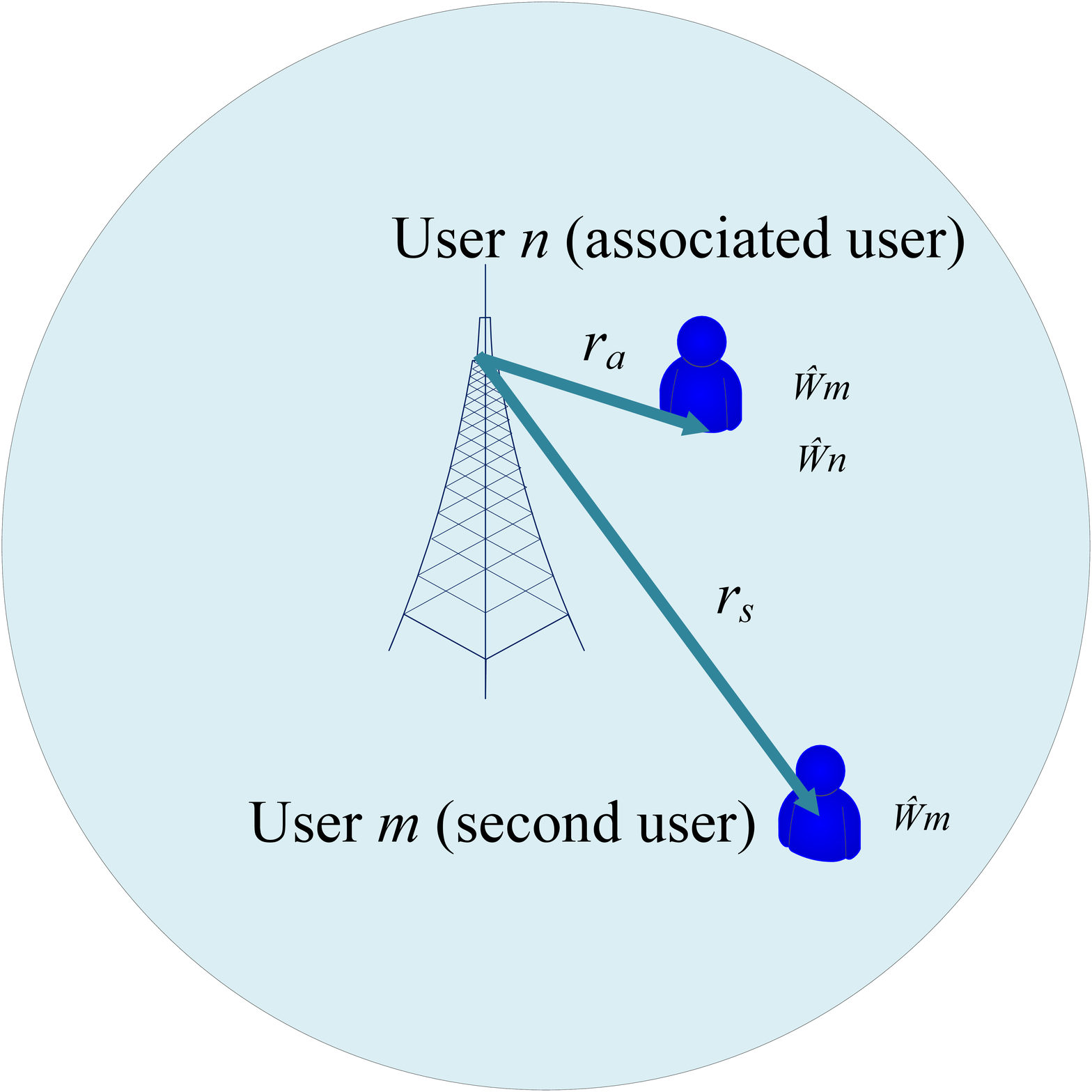}}
	\caption{NOMA senario cases.}
	\label{Fig.fig2}
	\vspace{-0.5cm}
\end{figure}

%In practice, multiple users are connected to the transmitter (UAV) one by one
Employing NOMA technique inspired from \cite{HetnetNOMAYZMAJ2017}, the first user is assumed to associate with BS in the previous round of the user association and the second user is also connected to the same BS. In existing works on NOMA-HetNet systems, the distance between the first connected NOMA user and its associated BS was fixed \cite{HetnetNOMAYZMAJ2017}. On the contrary, we consider this distance as a random variable $r_{a}$. The distance between the second user and the connected BS is also a random variable $r_{s}$.  The distances $r_{s}$ and $r_{a}$ follow the distribution ${f_{R_u}(r)}$ in (\ref{eq:04}) where $u\in\{a,s\}$. As shown in Fig. \ref{Fig.fig2}, two cases can be considered due to the uncertainty of whether the second user is farther or nearer. In Case \RNum{1}, user $m$ is the associated user and user $n$ is the second user which is nearer to  ${\text {BS}}_{k,q}$ ($r_{s}\le r_{a}$). In Case \RNum{2}, user  $n$ is the associated user and user  $m$ is the second user which is farther from ${\text {BS}}_{k,q}$ ($r_{s}>r_{a}$). Hence, user $m$ is always the far user and user $n$ is always the near user. The near user decodes messages of both $m$-th and $n$-th users and  performs SIC, while the far user only decodes its own message.
Thus, the following SINR is obtained for user $m$ by substituting $\{i\in m\}$  in  (\ref{eq:01}):
\begin{align}\label{eq:05}
&\gamma_{{m,m}_{k,q}}=\frac{{a_{m_{k,q}}}P_k{g_{m_{k,q}}}{d_{m_{k,q}}}^{-\alpha_k}}{{a_{n_{k,q}}P_k{g_{m_{k,q}}}{d_{m_{k,q}}}^{-\alpha_k}}+{I_{m_{k,q}}}+{\sigma^2}},
\end{align} 
where $I_{m_{k,q}}={\sum_{j=1}^K \hspace{0.3cm}\sum_{l\in\Phi_{b_j}\backslash\left \{BS_{k,q} \right \}}{P_j{g_{m_{j,l}}}d_{m{j,l}}^{-\alpha_j}}}$ is the interference at  user $m$ from the BSs in  all macrocells and smallcells  except its  serving BS.
Substituting $\{i\in n\}$  in  (\ref{eq:01}), the SINR at the $n$-th user decoding the message of user $m$ is: 
\begin{align}\label{eq:06}
&\gamma_{{n,m}_{k,q}}=\frac{{a_{m_{k,q}}}P_k{g_{n_{k,q}}}{d_{n_{k,q}}}^{-\alpha_k}}{{a_{n_{k,q}}P_k{g_{n_{k,q}}}{d_{n_{k,q}}}^{-\alpha_k}}+{I_{n_{k,q}}}+{\sigma^2}},
\end{align} 
where $I_{n_{k,q}}={\sum_{j=1}^K \hspace{0.3cm}\sum_{l\in\Phi_{b_j}\backslash\left \{BS_{k,q} \right \}}{P_j{g_{n_{j,l}}}d_{n{j,l}}^{-\alpha_j}}}$ is the interference at  user $n$ from BSs in all  macrocells and smallcells  except its serving BS. The SINR at user $n$ for decoding its own message (after SIC) is calculated as follows by substituting $\{i\in n\}$ in  (\ref{eq:01}),
\begin{align}\label{eq:07}
&\gamma_{{n},n_{k,q}}=\frac{{a_{n_{k,q}}}P_k{g_{n_{k,q}}}{d_{n_{k,q}}}^{-\alpha_k}}{{I_{n_{k,q}}}+{\sigma^2}}.
\end{align} 
The SINR at users for Case \RNum{1}  and Case \RNum{2}, respectively, can be obtained by substituting $\{{d_{m_{k,q}}}=r_{a},{d_{n_{k,q}}}=r_{s}\}$ and $\{{d_{m_{k,q}}}=r_{s},{d_{n_{k,q}}}=r_{a}\}$ in (\ref{eq:05}), (\ref{eq:06}), and (\ref{eq:07}). 
%\begin{flushright}
%	$\Box \; $
%	\end{flushright}

Considering non-colluding eavesdropping scenario, the received SINR at the most detrimental eavesdropper for detecting the message of the $w$-th user is obtained by substituting $\{i\in e \}$ in (\ref{eq:01}) as: 
\begin{align}\label{eq:08}
&{\gamma_{e_{max_{k,q}}}}=\max_{e\in\Phi_{e}}\frac{{a_{w_{k,q}}}P_k{g_{e_{k,q}}}d_{e_{k,q}}^{-\alpha_k}}{{I_{e_{k,q}}}+{\sigma^2}},
\end{align} 
where   $I_{e_{k,q}}={\sum_{j=1}^K \hspace{0.3cm}\sum_{l\in\Phi_{b_j}\backslash\left \{BS_{k,q} \right \}}{P_j{g_{e_{j,l}}}d_{e_{j,l}}^{-\alpha_j}}}$ is the interference at eavesdropper $e$ from BSs in all  macrocells and smallcells except ${\text {BS}}_{k,q}$. 

\subsection{Interference Characteristic Function}\label{IA}
	The characteristic function of interference from the  macro and small BSs at user $w$ is computed using \cite[Lemma~2]{HetnetNOMAYZMAJ2017} as:
	\begin{align}\label{eq:09}
	L_{I_{w_{k,q}}}(s)&=\E[e^{-sI_{w_{k,q}}}]=\exp\Big\{-\sum_{j=1}^{K}2\pi{\lambda_{j}}\frac{sP_j{(y_{j}(r_u))}^{2-\alpha_j}}{{\alpha_j}-2}\nonumber\\
	&\times{}_2\!F_1\Big(1,1-\frac{2}{\alpha_j};2-\frac{2}{\alpha_j};-\frac{sP_j}{ {(y_{j}(r_u))}^{\alpha_j}}\Big)\Big\},    
    \end{align}    
where $w\in\{m,n\}$, $u\in\{a,s\}$, and ${}_2\!F_1$ denotes the Gauss hypergeometric function \cite{BookII2007}.  The distance between user $w$ and the closest interferer in tier $j$ is  $y_{j}(r_u)={(\hat{P_j}\hat{B_j})^{1/\alpha_j}}{r_u^{1/\hat\alpha_j}}$.

	The characteristic function of interference at an eavesdropper is also obtained using \cite[Theorem~2]{HetnetSECLKMAS2016} as:
	\begin{align}\label{eq:10} 
    &L_{I_{e_{k,q}}}(s)=\E[e^{-sI_{e_{k,q}}}]=\nonumber\\
	&\exp\Big\{-\sum_{j=1}^{K}\pi{\lambda_{j}}{(sP_j)^{2/\alpha_j}}\Gamma\Big(1+\frac{2}{\alpha_j}\Big)\Gamma\Big(1-\frac{2}{\alpha_j}\Big)\Big\},
	\end{align}
where $\Gamma$ stands for the Gamma function \cite{BookII2007}.  For the sake of simplicity, $L_{I_{w_{k,q}}}$ and $L_{I_{e_{k,q}}}$ are shown as $L_{I_w}$ and $L_{I_e}$, respectively, in subsequent descriptions.

\section{Performance Analysis}
In this section, performance of the considered SN-Het is analyzed in term of the ergodic secrecy rate. To this end, we first study each tier of the SN-Het. Initially, the ergodic rates of the legitimate users are derived in Lemma \ref{le1}. The proof is in Appendix A.

\begin{lemma}\label{le1}
	Based on the location of NOMA users, the ergodic rate is presented below.

		1) Case \RNum{1}: The  ergodic rates of the associated user (user $m$) and the second user (user $n$) in the $k$-th tier are derived as (\ref{eq:11}) and (\ref{eq:12}), respectively. 
%	\begin{figure*}[!b]
	%	\hrulefill
		%\small
	\begin{align}\label{eq:11}
{R_{m,k}^{\textup{\RNum{1}}}}=&\frac{1}{\ln2}\int_{0}^{\infty}\int_{0}^{r_{a}}\int_{0}^{\frac{a_{m_{k,q}}}{a_{n_{k,q}}}}{\frac{1}{1+t}}{\exp\Big\{-\frac{t{\sigma^2}({r_{a}}^{\alpha_k}+{r_{s}}^{\alpha_k})}{AP_k}\Big\}}\nonumber\\
&{{L_{I_m}}\Big\{\frac{{t}{r_{a}}^{\alpha_k}}{AP_k}\Big\}}{{L_{I_n}}\Big\{\frac{{t}{r_{s}}^{\alpha_k}}{AP_k}\Big\}}{f_{R_s}(r_{s})}{f_{R_a}(r_{a})}\,dt\,dr_s\,dr_a,
	\end{align} 
%\normalsize
	\begin{align}\label{eq:12}
	{R_{n,k}^{\textup{\RNum{1}}}}=&\frac{1}{\ln2}\int_{0}^{\infty}\int_{0}^{\infty}{\frac{1}{1+t}}{\exp\Big\{-\frac{t{\sigma^2}{r_{s}}^{\alpha_k}}{{a_{n_{k,q}}}P_k}\Big\}}{{L_{I_n}}\Big\{\frac{{t}{r_{s}}^{\alpha_k}}{{a_{n_{k,q}}}P_k}\Big\}}\nonumber\\
	&{(1-F_{R_a}(r_{s}))}{f_{R_s}(r_{s})}\,dt\,dr_s.
	\end{align}
%	\end{figure*}

	2) Case \RNum{2}: The ergodic rates of the associated user (user $n$) and the second user (user $m$) in the $k$-th tier are derived as (\ref{eq:13}) and (\ref{eq:14}), respectively.
	%\begin{figure*}[!t]
	%	\begin{figure*}[!b]
	\begin{align}\label{eq:13}
	{R_{n,k}^{\textup{\RNum{2}}}}=&\frac{1}{\ln2}\int_{0}^{\infty}\int_{0}^{\infty}{\frac{1}{1+t}}{\exp\Big\{-\frac{t{\sigma^2}{r_{a}}^{\alpha_k}}{{a_{n_{k,q}}}P_k}\Big\}}{{L_{I_n}}\Big\{\frac{{t}{r_{a}}^{\alpha_k}}{{a_{n_{k,q}}}P_k}\Big\}}\nonumber\\
	&{(1-F_{R_s}(r_{a}))}{f_{R_a}(r_{a})}\,dt\,dr_a,
	\end{align} 
	\begin{align}\label{eq:14}
	{R_{m,k}^{\textup{\RNum{2}}}}=&\frac{1}{\ln2}\int_{0}^{\infty}\int_{r_{a}}^{\infty}\int_{0}^{\frac{a_{m_{k,q}}}{a_{n_{k,q}}}}{\frac{1}{1+t}}{\exp\Big\{-\frac{t{\sigma^2}({r_{s}}^{\alpha_k}+{r_{a}}^{\alpha_k})}{AP_k}\Big\}}\nonumber\\
	&{{L_{I_m}}\Big\{\frac{{t}{r_{s}}^{\alpha_k}}{AP_k}\Big\}}{{L_{I_n}}\Big\{\frac{{t}{r_{a}}^{\alpha_k}}{AP_k}\Big\}}{f_{R_s}(r_{s})}{f_{R_a}(r_{a})}\,dt\,dr_s\,dr_a.	
	\end{align}
where $A\triangleq\frac{1}{{a_{m_{k,q}}}-t{a_{n_{k,q}}}}$.
\end{lemma} 
 It should be noted that due to higher limit of the internal integral in (\ref{eq:11}) and (\ref{eq:14}), the integrand is easy to compute. Despite non-closed-form equations for ergodic rates of the legitimate users, they are efficiently computed numerically compared to Monte Carlo simulations which depend on repeated random sampling.
 \begin{remark}\label{Re1}
 	 Considering the incremental behavior of distribution ${f_{R_u}(r)}$  with growth in density of BSs at the $k$-th tier ($\lambda_{k}$) in Lemma \ref{le1} and also increase in the exponential term and the characteristic function of interference with enhanced transmission power of BSs at the $k$-th tier ($P_{k}$), the ergodic rate of $k$-th tier is incremental.  
 	 %Furthermore, it is confirmed that the ergodic rate in the other tiers is also increased due to the efficient association of cell edge users in these tiers with the $k$-th tier.
 	 \end{remark}
 	 	 To present simpler expressions, in the following lemma, lower bounds on the results of Lemma \ref{le1} are derived. The proof is in Appendix B.

\begin{lemma}\label{le2}
	The lower bounds on the ergodic rates of the associated and the second users in the $k$-th tier for Case \RNum{1} and Case \RNum{2} are obtained at the top of this page for $ w\in\{m,n\}$, we have $\E[{I_{w_{k,q}}}]={\sum_{j=1}^K \hspace{0.3cm}(\frac{2\pi P_j{\lambda_{j}}}{\alpha_j-2}){y_{j}(r_u)}^{2-\alpha_j}}$. Note that all logarithms are in base 2 in this paper. 
	\begin{figure*}[!t] 
		  \vspace{-0.4cm}
	\begin{align}\label{eq:15}
	{\bar{R}_{m,k}^{\text{\RNum{1}}}}&=\log\Big(1+\Big(\frac{2{a_{n_{k,q}}}}{{a_{m_{k,q}}}}+{\frac{1}{{{a_{m_{k,q}}}{P_k}}}}\Big(\int_{0}^{\infty}\big\{\E[{I_{n_{k,q}}}]+{\sigma^2}\big\}{{r_s}^{\alpha_k}}(1-{F_{R_a}(r_s)}){f_{R_s}(r_s)}\,dr_s\nonumber\\
	&+\int_{0}^{\infty}\big\{\E[{I_{m_{k,q}}}]+{\sigma^2}\big\}{{r_a}^{\alpha_k}}{f_{R_a}(r_a)}\,dr_a\Big)\Big)^{-1}\Big),
	\end{align}
	\begin{align}\label{eq:16}
	{\bar{R}_{n,k}^{\text{\RNum{1}}}}=\log\Big(1+\Big({\frac{1}{{{a_{n_{k,q}}}{P_k}}}\int_{0}^{\infty}\big\{\E[{I_{n_{k,q}}}]+{\sigma^2}\big\}{{r_s}^{\alpha_k}}(1-{F_{R_a}(r_s)}){f_{R_s}(r_s)}\,dr_s}\Big)^{-1}\Big),
	\end{align}
	\begin{align}\label{eq:17}
		{\bar{R}_{n,k}^{\text{\RNum{2}}}}=\log\Big(1+\Big({\frac{1}{{{a_{n_{k,q}}}{P_k}}}\int_{0}^{\infty}\big\{\E[{I_{n_{k,q}}}]+{\sigma^2}\big\}{{r_a}^{\alpha_k}}{f_{R_a}(r_a)}\,dr_a}\Big)^{-1}\Big),
		\end{align}
		\begin{align}\label{eq:18}
		{\bar{R}_{m,k}^{\text{\RNum{2}}}}&=\log\Big(1+\Big(\frac{2{a_{n_{k,q}}}}{{a_{m_{k,q}}}}+{\frac{1}{{{a_{m_{k,q}}}{P_k}}}}\Big(\int_{0}^{\infty}\big\{\E[{I_{m_{k,q}}}]+{\sigma^2}\big\}{{r_s}^{\alpha_k}}{F_{R_a}(r_s)}{f_{R_s}(r_s)}\,dr_s\nonumber\\
		&+\int_{0}^{\infty}\big\{\E[{I_{n_{k,q}}}]+{\sigma^2}\big\}{{r_a}^{\alpha_k}}{f_{R_a}(r_a)}\,dr_a\Big)\Big)^{-1}\Big).
	\end{align} 
		\hrulefill
\end{figure*}
\end{lemma}

The derived expressions in Lemma \ref{le2},  relaxe the ergodic rates of the legitimate users in the $k$-th tier of Lemma \ref{le1} to a single integral form. Now, we consider a special case where the  path loss exponents for all tiers are equal and the association is unbiased and we derive the rate expressions in closed-forms. The proof is in Appendix C.
\begin{corollary}\label{co1}
Assuming the same path loss exponent, $\left\{\alpha_k\right\}_{k=1,...,K}=\alpha$ for all tiers and an unbiased association, $\hat{B_j}=1$, the lower bounds on the ergodic rates of the associated and the second users in the $k$-th tier for Case \RNum{1} and Case \RNum{2}  are expressed in closed-form as:
\begin{align}\label{eq:19}
{\tilde{R}_{m,k}^{\textup{\RNum{1}}}}&=\log\Big(1+\Big(\frac{2{a_{n_{k,q}}}}{{a_{m_{k,q}}}}+\frac{1}{{{a_{m_{k,q}}}}}\Big({\frac{5}{2(\alpha-2)}}\nonumber\\
&+{\frac{\tilde{A}(1+2^{\frac{\alpha}{2}+1})}{2}}\Big)\Big)^{-1}\Big),
\end{align}
\begin{align}\label{eq:20}
{\tilde{R}_{n,k}^{\textup{\RNum{1}}}}=\log\Big(1+\Big(\frac{1}{a_{n_{k,q}}}\Big({\frac{1}{2(\alpha-2)}}+{\frac{\tilde{A}}{{2}}}\Big)\Big)^{-1}\Big),
\end{align}
\begin{align}\label{eq:21}
{\tilde{R}_{n,k}^{\textup{\RNum{2}}}}=\log\Big(1+\Big(\frac{1}{{a_{n_{k,q}}}}\Big({\frac{2}{\alpha-2}}+{\tilde{A}2^{\frac{\alpha}{2}}}\Big)\Big)^{-1}\Big),
\end{align}
\begin{align}\label{eq:22}
{\tilde{R}_{m,k}^{\textup{\RNum{2}}}}&=\log\Big(1+\Big(\frac{2{a_{n_{k,q}}}}{{a_{m_{k,q}}}}+\frac{1}{{{a_{m_{k,q}}}}}\Big({\frac{7}{2(\alpha-2)}}\nonumber\\
&+{\frac{\tilde{A}(2^{{\frac{\alpha}{2}}+2}-1)}{2}}\Big)\Big)^{-1}\Big),
\end{align} 

where $\tilde{A}\triangleq\frac{{\sigma^2}\Gamma({\frac{\alpha}{2}+1})}{P_k(2{E})^{\frac{\alpha}{2}}}$ and ${E}\triangleq\sum_{j=1}^K{\pi{\lambda_j}{P_j}^{\frac{2}{\alpha}}}$.

%If $\alpha=4$, the lower bound on the ergodic rate of the users  in $k$-th tier are given as
%\begin{align}\label{eq:23}
%{\tilde{R}_{n,k,L}^{\text{Case\RNum{1}}}}\mid_{\alpha=4}=\log\Big(1+\Big(\frac{\Theta}{a_{n_{k,q}}}\Big({\frac{1}{4}}+{\frac{{\sigma^2}}{4{P_k}(\sum_{j=1}^K{{\lambda_j}{{P_j}}^{\frac{1}{2}}})^{2}}}\Big)\Big)^{-1}\Big),
%\end{align}
%\begin{align}\label{eq:24}
%{\tilde{R}_{n,k,L}^{\text{Case\RNum{2}}}}\mid_{\alpha=4}=\log\Big(1+\Big(\frac{\Theta}{{a_{n_{k,q}}}}\Big({1}+{\frac{2{\sigma^2}}{(\sum_{j=1}^K{{\lambda_j}{{P_j}}^{\frac{1}{2}}})^2}}\Big)\Big)^{-1}\Big),
%\end{align}
%and 
%\begin{align}\label{eq:25}
%&{\tilde{R}_{m,k,L}^{\text{Case\RNum{1}}}}\mid_{\alpha=4}=\log\Big(1+\Big(\frac{2{a_{n_{k,q}}}}{{a_{m_{k,q}}}}+\frac{\Theta}{{{a_{m_{k,q}}}}}\Big(\frac{{5}}{4}+{\frac{9{\sigma^2}}{4P_k(\sum_{j=1}^K{{\lambda_j}{{P_j}}^{\frac{1}{2}}})^{2}}}\Big)\Big)^{-1}\Big),
%\end{align}
%\begin{align}\label{eq:26}
%&{\tilde{R}_{m,k,L}^{\text{Case\RNum{2}}}}\mid_{\alpha=4}=\log\Big(1+\Big(\frac{2{a_{n_{k,q}}}}{{a_{m_{k,q}}}}+\frac{\Theta}{{{a_{m_{k,q}}}}}\Big({\frac{7}{4}}+{\frac{15{\sigma^2}}{4P_k(\sum_{j=1}^K{{\lambda_j}{{P_j}}^{\frac{1}{2}}})^{2}}}\Big)\Big)^{-1}\Big),
%\end{align} 

\end{corollary}

Next, the ergodic leakage rate of the most detrimental eavesdropper is provided in Lemma \ref{le3}. The proof is in Appendix D.
\begin{lemma}\label{le3}
For $w\in\{m,n\}$, the ergodic leakage rate at the most detrimental eavesdropper for decoding the message of $w$-th user in the $k$-th tier is expressed as:
		\begin{align}\label{eq:23}
	{R_{e,k}^{w}}&=\frac{1}{\ln2}\int_{0}^{\infty}{\frac{1}{1+t}}\Big(1-\exp\Big\{\int_{0}^{\infty}-2\pi{\lambda_e}\nonumber\\
	&\times{}{\exp\Big\{-\frac{t{\sigma^2}{r}^{\alpha_k}}{{a_{w_{k,q}}}P_k}\Big\}}{{L_{I_e}}\Big\{\frac{{t}{r}^{\alpha_k}}{{a_{w_{k,q}}}P_k}\Big\}}r\,dr\Big\}\Big)\,dt.
		\end{align} 
\end{lemma}
\begin{remark}\label{Re2}
Eq. (\ref{eq:23}) confirms that higher density of eavesdroppers in the network leads to the ergodic leakage rate increase. 
\end{remark}
Now, the special case of interference-limited network ($\sigma^2=0$) with unbiased association and equal path loss exponents for all tiers is investigated. The interference-limited regime is of high importance in HetNets due to high BS density  which results in domination of  interference to the noise power \cite{HetnetHYPJ2012}. In the following corollary, the results of Lemma \ref{le1} and Lemma \ref{le3} are further simplified for this special case. The proof is in Appendix E.
\begin{corollary}\label{co2}
	The ergodic rates of the associated and the second users in the $k$-th tier of Case \RNum{1} and Case \RNum{2} are presented below considering some special conditions: (i) the same path loss exponent for all tiers ($\left\{\alpha_k\right\}_{k=1,...,K}=\alpha$), (ii) an unbiased association ($\hat{B_j}=1$), and (iii) the interference-limited regime ($\sigma^2=0$).
	\begin{align}\label{eq:24}
	{R_{m,k}^{\textup{\RNum{1}}}(\alpha,1)}&={R_{m,k}^{\textup{\RNum{2}}}(\alpha,1)}\nonumber\\
	&=\frac{1}{\ln2}\int_{0}^{\frac{a_{m_{k,q}}}{a_{n_{k,q}}}}{\frac{1}{2(1+t)(1+Z_m)^2}}\,dt,
	\end{align} 
	\begin{align}\label{eq:25}
	{R_{n,k}^{\textup{\RNum{1}}}(\alpha,1)}={R_{n,k}^{\textup{\RNum{2}}}(\alpha,1)}=\frac{1}{\ln2}\int_{0}^{\infty}{\frac{1}{(1+t)(2+Z_n)}}\,dt,
	\end{align}
		where
	\begin{align}\label{eq:26}
		Z_m&\triangleq\frac{2tA}{(\alpha-2)}\times{}_2\!F_1\Big(1,1-\frac{2}{\alpha};2-\frac{2}{\alpha};-{t}{A}\Big),
	\end{align} 
	\begin{align}\label{eq:27}
	&Z_n\triangleq\frac{2t}{{a_{n_{k,q}}}(\alpha-2)}\times{}{}_2\!F_1\Big(1,1-\frac{2}{\alpha};2-\frac{2}{\alpha};-\frac{t}{{a_{n_{k,q}}}}\Big).
	\end{align}

	In addition, the ergodic leakage rate at the most detrimental eavesdropper for detecting the information of  $w$-th user ($w\in\{m,n\}$) in the $k$-th tier is given by:
	\begin{align}\label{eq:28}
	{R_{e,k}^{w}(\alpha,1)}=\frac{1}{\ln2}\int_{0}^{\infty}{\frac{1}{1+t}}\Big(1-\exp\Big\{-{B}{t^{-\frac{2}{\alpha}}}\Big\}\Big)\,dt,
	\end{align}
	where
	\begin{align}\label{eq:29}
	B\triangleq\lambda_e{a_{w_{k,q}}}^{\frac{2}{\alpha}}\Big(\sum_{j=1}^{K}\lambda_{j}\hat{P_j}^{\frac{2}{\alpha}}\Gamma\Big(1+\frac{2}{\alpha}\Big)\Gamma\Big(1-\frac{2}{\alpha}\Big)\Big)^{-1}.
	\end{align}
\end{corollary}
 The provided ergodic rates of the legitimate users at (\ref{eq:24}) and (\ref{eq:25}) in the interference-limited regime neither depend on the number of tiers, nor on BS transmit power and BS density. This is consistent with the result of \cite{HetnetHYPJ2012,andrews2011tractable}.

After studying each tier in the considered SN-Het system model, the ergodic secrecy rate of the SN-Het is calculated in  Theorem \ref{Th1} below.

\begin{theorem}\label{Th1}
	The ergodic secrecy rate of the SN-Het is calculated as follows:
	\begin{align}\label{eq:30}
	&\mathcal{R}_{sec}=\sum_{k=1}^K{A_{k}(R_{sec,k}^{\textup{\RNum{1}}}+R_{sec,k}^{\textup{\RNum{2}}})},\\
	&R_{sec,k}^{\textup{\RNum{1}}}={\left[{R_{n,k}^{\textup{\RNum{1}}}}-{R_{e,k}^{n}}\right]^{+}}+{\left[{R_{m,k}^{\textup{\RNum{1}}}}-{R_{e,k}^{m}}\right]^{+}},\nonumber\\
	&R_{sec,k}^{\textup{\RNum{2}}}={\left[{R_{n,k}^{\textup{\RNum{2}}}}-{R_{e,k}^{n}}\right]^{+}}+{\left[{R_{m,k}^{\textup{\RNum{2}}}}-{R_{e,k}^{m}}\right]^{+}},\nonumber
	\end{align} 
    where $\left[x\right]^{+}=\max\left\{x,0\right\}$, $R_{sec,k}^{\textup{\RNum{1}}}$ and $R_{sec,k}^{\textup{\RNum{2}}}$ represent the ergodic secrecy rates in the $k$-th tier for Case \RNum{1} and Case~ \RNum{2}, respectively. ${\left[{R_{n,k}^{\text{\RNum{1}}}}-{R_{e,k}^{n}}\right]^{+}}$ and ${\left[{R_{m,k}^{\text{\RNum{1}}}}-{R_{e,k}^{m}}\right]^{+}}$ denote the ergodic secrecy rates of the users $n$ and $m$ in the $k$-th tier for Case \RNum{1}, respectively (similarly for Case \RNum{2}). $A_{k}$ is given in (\ref{eq:03}), ${R_{m,k}^{\textup{\RNum{1}}}}$, ${R_{n,k}^{\textup{\RNum{1}}}}$, ${R_{n,k}^{\textup{\RNum{2}}}}$, ${R_{m,k}^{\textup{\RNum{2}}}}$, $R_{e,k}^{n}$, and $R_{e,k}^{m}$ are also derived in (\ref{eq:11})-(\ref{eq:14}) and  (\ref{eq:23}).  
    \begin{proof}
    Based on \cite[Proposition~1]{HetnetNOMAYZMAJ2017} at non-secure mode, the achievability  of (\ref{eq:30}) could be shown by using  wiretap coding for each of NOMA users. $R_{sec,k}^{\textup{\RNum{1}}}$ and $R_{sec,k}^{\textup{\RNum{2}}}$ are inferred from  Lemma \ref{le1} and Lemma \ref{le3}.
    \end{proof}
\end{theorem}
\begin{remark}\label{Re3}
  Based on Theorem \ref{Th1}, a lower bound on the ergodic secrecy rate ($\mathcal{\bar{R}}_{sec}$) is obtained by substituting  (\ref{eq:15})-(\ref{eq:18})  (i.e., the lower bounds on ergodic rates of the legitimate users (${\bar{R}_{w,k}}$)) into (\ref{eq:30}). 
\end{remark}

\begin{remark}\label{Re4}
The ergodic secrecy rate of the SN-Het is derived for the special case (the same per-tier path loss exponent, interference-limited network, and unbiased association) by applying derived expressions in Corollary \ref{co2} to Theorem \ref{Th1}. 
\end{remark}

\begin{remark}\label{Re5}
	 Assuming one tier in the SN-Het ($K=1$), the ergodic secrecy rate of multi-cell NOMA system is obtained by invoking the results of Theorem \ref{Th1}. To the best of our knowledge, it is not addressed yet in the literature. 
	\end{remark}

%\subsection{Energy efficiency}
 % In this section, we proceed to study the energy efficiency metric in the proposed secure NOMA-based HetNet. similar to \cite{Ref1}, the energy efficiency for $k$-th tier is evaluated as
 %\begin{align}\label{eq:26}
%\mathrm{EE}_k=\frac{R_{sec,k}^{\text{near}}+R_{sec,k}^{\text{far}}}{P_k^{\text{static}}+\frac{P_k}{\epsilon_k}},
 % \end{align}
  %where $R_{sec,k}^{\text{near}}+R_{sec,k}^{\text{far}}$ are given by (\ref{eq:17}), $P_k^{\text{static}}$ is  the static hardware power consumption of macrocell and smallcells BSs in $k$-th tier and $\epsilon_k$ is the efficiency of the power amplifier.
  
  %Considering all tiers, the energy efficiency of the proposed secure NOMA-based HetNet is 
  %\begin{align}\label{eq:27}
  %\mathrm{EE}^{\text{total}}=\sum_{k=1}^K{A_{k}\mathrm{EE}_k}.
  %\end{align}
  %where $A_{k}$ is given by (\ref{eq:03}).
  
  \section{Numerical Results}
  The derived analytical results ($\mathcal{R}_{sec}$ in (\ref{eq:30})) are numerically evaluated in this section. The Monte Carlo simulation results are also provided to verify the obtained analytical results. The general parameters used in the analytical derivations and simulations are summarized in  Table \ref{table:1}, which are consistent with the papers \cite{HetnetHYPJ2012,HetnetNOMAYZMAJ2017,HetnetSECLKMAS2016}. The Monte Carlo simulation area is a circle with radius  $10^4$ m.
  \begin{table}[h!]
  	\centering
  	%\vspace{-0.3cm}
  	\caption{Network Parameters }
  	\vspace{-0.2cm}
  	\renewcommand{\arraystretch}{1.1}
  	%|c|p{2.5cm}|
  	
  	\begin{tabular} {|>{\centering}p{0.2\textwidth} |>{\centering\arraybackslash}p{0.2\textwidth}|} 
  		\hline
  		Parameter &   Value
  		\\
  		\hline
         Number of iterations for simulations  &  $10^6$ 
  		\\
  		\hline
  		 AWGN power  & $\sigma^2=-90$ dBm
  		\\
  		\hline
  		 BS transmit power & $P_1=$40 dBm, $P_2=$30 dBm, $P_3=$20 dBm
  		\\  
  		\hline
  		Path loss exponents & $\alpha_1=$3.5, $\alpha_2=\alpha_3=$4
  		\\  
  		\hline
  		 NOMA power sharing coefficients\tablefootnote{The NOMA power sharing coefficients for all tiers are to be same: $(a_{m_{k,q}},a_{n_{k,q}})=(a_m,a_n)$. Note that, for any ${a_m}$, we characterize the ergodic rate, and optimizing the ergodic rate versus ${a_m}$ is an interesting future work. However, for simulations, we choose ${1-a_{m}}<0.5$ to reduce the probability of SIC failing and thus achieving higher ergodic rates.}  &  $(0.6,0.4)$
  		\\  
  		\hline
  		 Bias factor of the first tier & $B_1=$1
  		\\  
  		\hline
  	\end{tabular}
  	\label{table:1}
	\vspace{-0.2cm}
  \end{table}

\begin{figure}
  	\centering
  \vspace{-0.5cm}
  	  	\includegraphics[width=6cm,height=5cm]{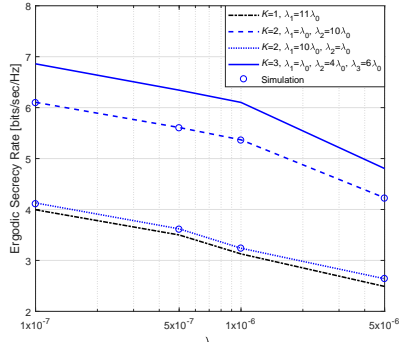}  
  	\vspace{-0.3cm}
  	\caption{The ergodic secrecy rate versus density of
  		eavesdroppers ($r_{a}=50, \lambda_0=\frac{1}{\pi500^2}=1.2732\times 10^{-6}, \{B_2, B_3\}=\{1, 1\}$).}
  	\label{Fig.fig3}
	\vspace{0.0cm}
  \end{figure}

    Fig. \ref{Fig.fig3} depicts behavior of the SN-Het ergodic secrecy rate versus $\lambda_e$ for different numbers of network tiers. $\lambda_e$ is drawn logarithmically due to its variable interval. The goal is to analyze the ergodic secrecy rate as number of tiers increase while a fixed density of BSs (i.e., $11{\lambda_0}$) is divided between tiers. As expected, the ergodic secrecy rate decreases as $\lambda_e$ increases, due to the increased leakage. Moreover, it is inferred that the ergodic rate improves as the number of tiers increases. It is worth noting that increase in the number of tiers reduces the inter-tier interference because of $P_1>P_2>P_3$ and $\alpha_2,\alpha_3>\alpha_1$. Interestingly, adding one tier is more beneficial for $K=1$ to $K=2$ compared with $K=2$ to $K=3$. In addition, another purpose in Fig. \ref{Fig.fig3} is to allocate a fixed density of BSs ($11{\lambda_0}$) between two tiers. It is observed that assigning more BSs to a tire with lower power and higher path loss ($K=2$) is more advantageous due to the reduction of interference. The analytical curves have a precise match to the results obtained with the Monte Carlo simulations. Due to the high computational complexity for Monte Carlo simulations, we only include the simulation results in  Fig. \ref{Fig.fig3}.
  
  %\begin{figure}
  %	\centering
  %	\vspace{-0.1cm}
  %	\includegraphics[width=0.5 \linewidth]{fig4.eps} 
  	%  	\includegraphics[width=6cm,height=5cm]{fig4.eps}  
  %	\vspace{-0.3cm}
  %	\caption{The ergodic secrecy rate versus density of
  %		eavesdroppers ($K=2, r_{a}=50,  \lambda_0=\frac{1}{\pi500^2}=1.2732\times 10^{-6}, B_2=1$).}
  %	\label{Fig.fig4}
  %	\vspace{-0.5cm}
%  \end{figure}
  
\begin{figure}
  	\centering
  	\vspace{-0.5cm}
  	\includegraphics[width=6cm,height=5cm]{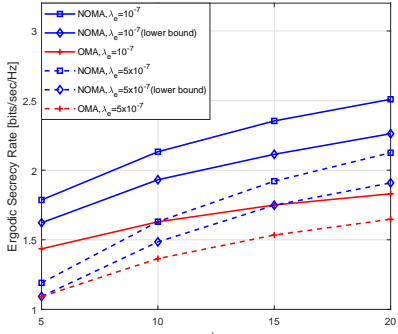}  
  	\vspace{-0.3cm}
  	\caption{The ergodic secrecy rate versus density of BSs in the second tier ($K=2,\lambda_1=\frac{1}{\pi500^2}=1.2732\times 10^{-6},  B_2=1$).}
  	\label{Fig.fig4}
	\vspace{-0.3cm}
  \end{figure}

   In Fig. \ref{Fig.fig4}, behavior of the two-tier SN-Het ergodic secrecy rate is demonstrated versus $\lambda_2$ for both NOMA and OMA strategies. Unlike the previous analyses and existing results in \cite{HetnetNOMAYZMAJ2017}, the first user distance ($r_a$) is not fixed here and the ergodic secrecy rate is computed accordingly. The obtained lower bound on ergodic secrecy rate according to Remark \ref{Re3}  is also plotted (denoted as lower bound). It is shown that NOMA significantly outperforms OMA. Furthermore, it is inferred that the ergodic secrecy rate improves as $\lambda_2$ increases. Not only the ergodic rate of the second tier is increased by increasing $\lambda_2$, but also the ergodic rate of the first tier increases, interestingly. This is due to the fact that increasing the second tier BSs results in more users with low SINR (i.e., at cell edge) at the first tier become associated with the second tier. In addition, increasing  $\lambda_2$ results in more interference at the eavesdroppers and degrades the eavesdroppers' channels. As it is inferred, the performance of NOMA is significantly better than OMA systems in networks with higher density of BSs which caused more interference. It is also observed that the obtained ergodic secrecy rates are reasonably close to the demonstrated lower bounds. 
   
\begin{figure}
 	\centering
 		\vspace{-0.5cm}
 	  	\includegraphics[width=6cm,height=5cm]{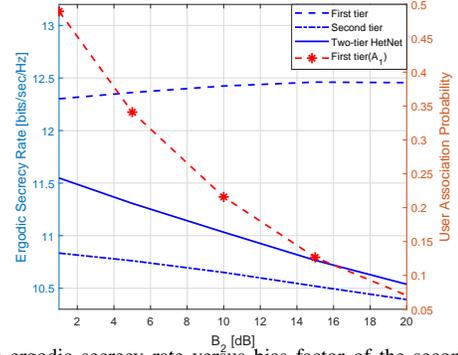} 
 	\vspace{-0.3cm}
 	\caption{The ergodic secrecy rate versus bias factor of the second tier ($K=2, r_{a}=10,  \lambda_1=\frac{1}{\pi500^2}=1.2732\times 10^{-6}, \lambda_2=15\lambda_1, \lambda_e=10^{-7}$).}
 	\label{Fig.fig5}
 	\vspace{-0.0cm}
 \end{figure}

The ergodic secrecy rate versus $B_2$ is presented in Fig. \ref{Fig.fig5} for all network tiers. As observed in \cite{HetnetHYPJ2012}, which is consistent with the prior existing works, the unbiased association always outperforms biasing from the study of rate at the overall network point of view. It is observed that increasing in $B_2$ causes the ergodic secrecy rate of the first tier to increase, while the ergodic secrecy rate decreases in the second tier. It is worth noting that more macro users with low SINR (i.e., at cell edge) are associated with the second tier as $B_2$ increases. This increased macro user association in the second tier degrades the ergodic secrecy rate of corresponding tier, but improves those of the first tier. However, reduction in the second tier ergodic secrecy rate is compensated by increase in  $\lambda_2$. Thus, the biased association is an efficient method in load balancing between each tier of the HetNet. Furthermore, the probability of association to the second tier increases as $B_2$ increases, which causes a faster drop in the two-tier case compared to the second tier case.

\begin{figure}
 	\centering
 		\vspace{-0.45cm}
 	\includegraphics[width=6cm,height=5cm]{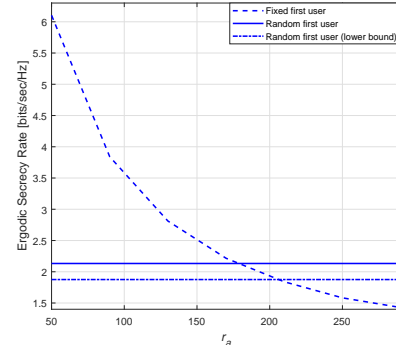}  
 	\vspace{-0.3cm}
 	\caption{The ergodic secrecy rate versus the first user distance ($K=2, \lambda_1=\frac{1}{\pi500^2}=1.2732\times 10^{-6}, \lambda_2=10\lambda_1, \lambda_e=10^{-7}$).}
 	\label{Fig.fig6}
	\vspace{-0.3cm}
 \end{figure}

Randomness to the first user distance is perceived in Fig. \ref{Fig.fig6} as one of our contributions contrary to \cite{HetnetNOMAYZMAJ2017}. It is shown by dashed line that the ergodic secrecy rate is highly depended on the value of $r_a$ with fixed first user location (as in \cite{HetnetNOMAYZMAJ2017}). Hence, the average of ergodic secrecy rate over the locations of the first user is also depicted by solid line. It is also inferred that obtained lower bound on the ergodic secrecy rate according to Remark \ref{Re3} (denoted as lower bound) is following the performance of ergodic secrecy rate.

 \subsection{Comparision}\label{IA}
 Since the foundations of the studied model are secrecy, NOMA, and HetNet, we compare the derived results with 1) the HetNet with secrecy constraint, 2) NOMA technique with secrecy constraint, and 3) the HetNet and NOMA-HetNet without secrecy constraints, as follows.

\begin{figure*}
	\centering
	\vspace{-0.5cm}
	\subfloat[]{\includegraphics[width=5.5cm]{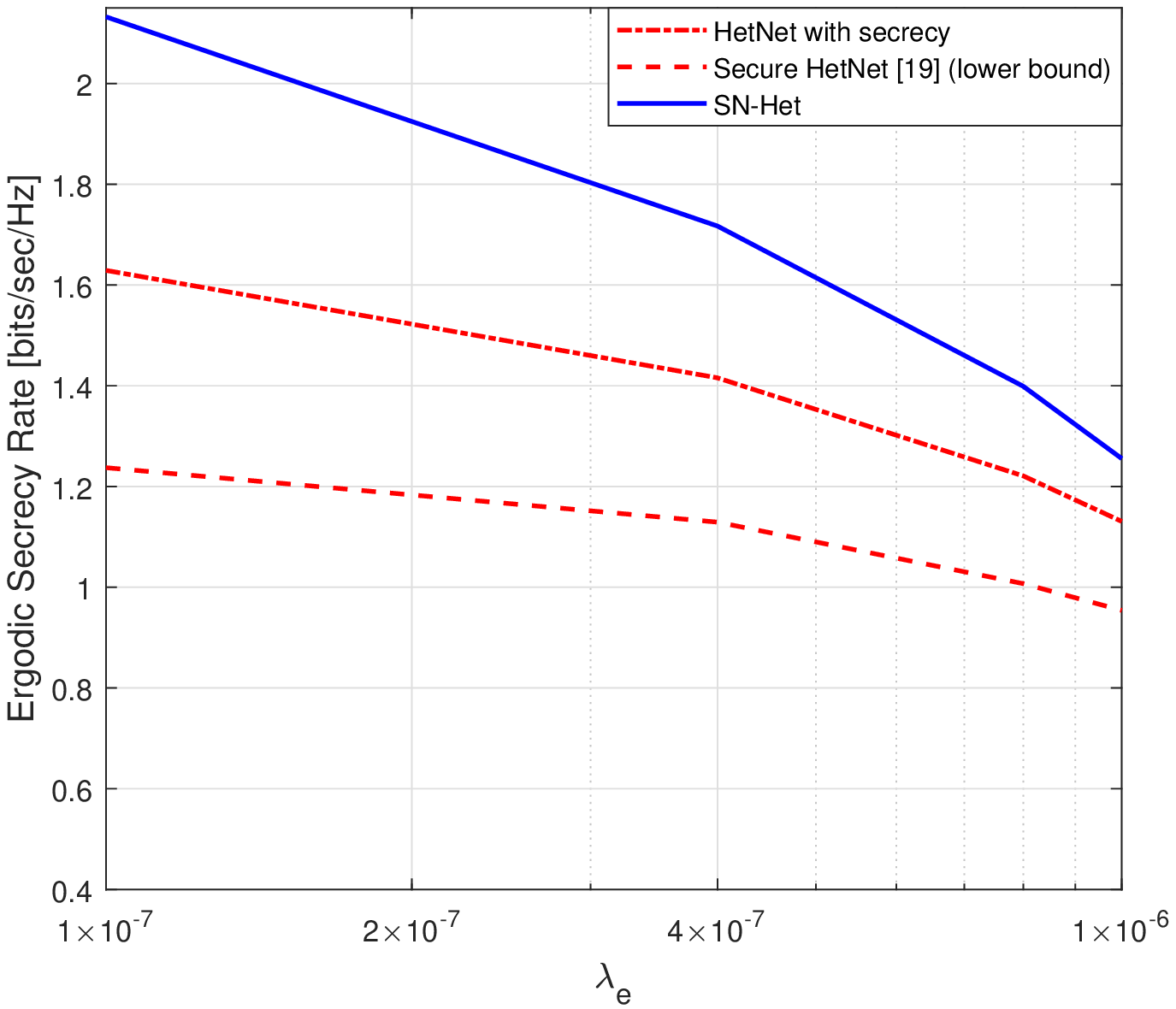}}
	\hspace{-0.0cm}
	\subfloat[]{\includegraphics[width=5.5cm]{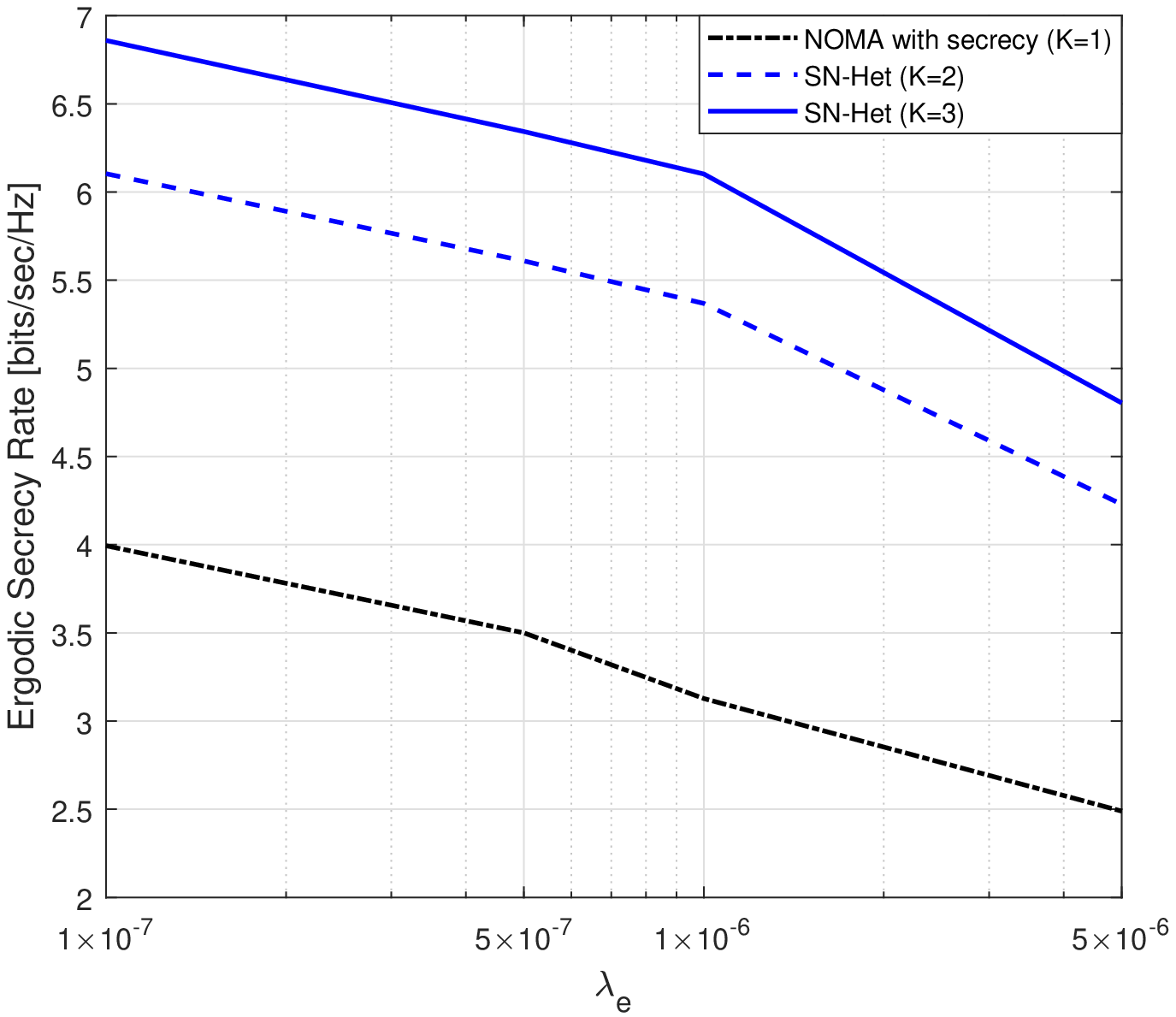}}
	\hspace{-0.0cm}
	\subfloat[]{\includegraphics[width=5.5cm]{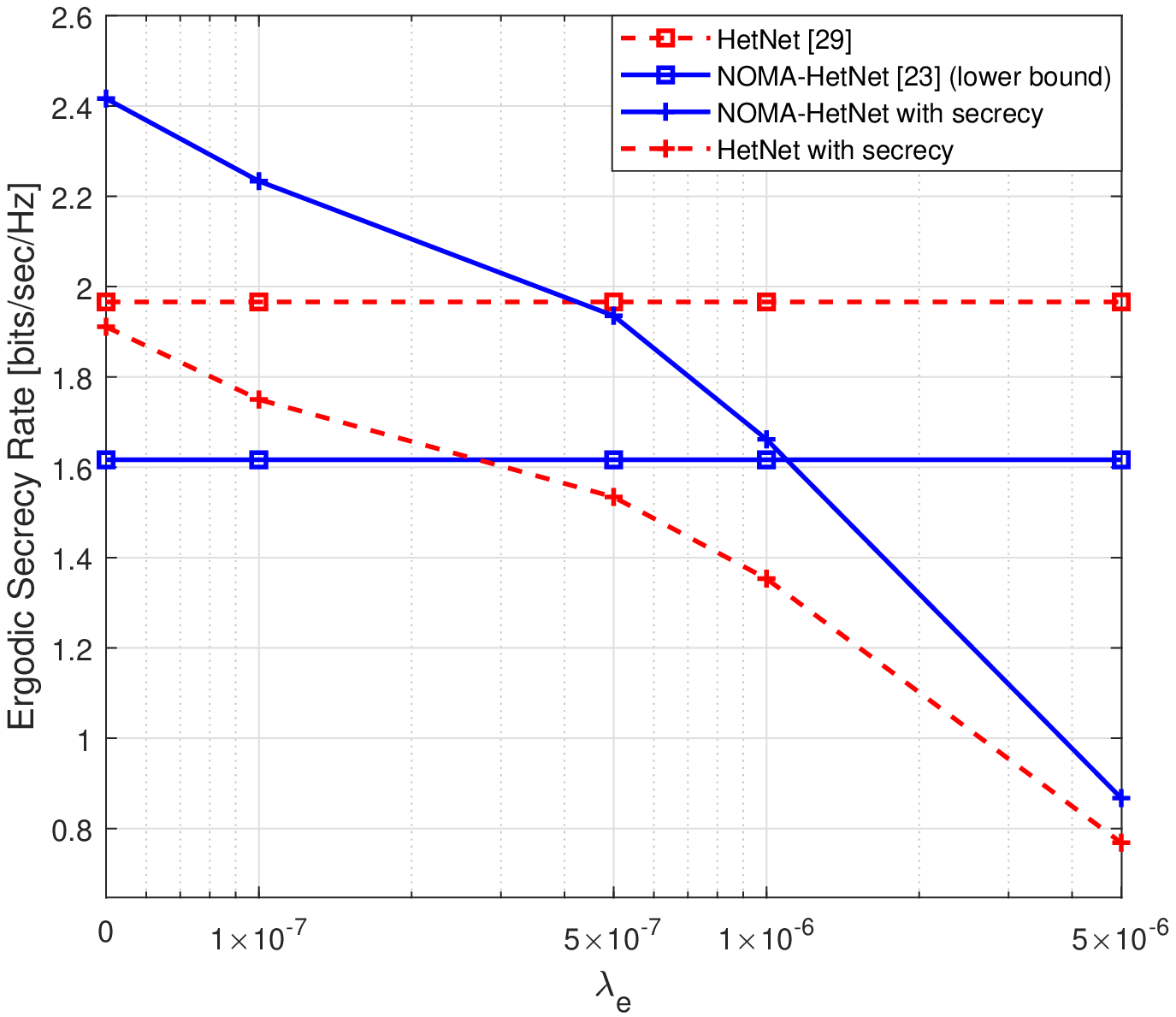}}
	\caption{The effects of (a) NOMA technique ($K=2,  \lambda_1=\frac{1}{\pi500^2}=1.2732\times 10^{-6}, \lambda_2=10\lambda_1, B_2=1$), (b) multi-tier using HetNet ($r_{a}=50, \lambda_0=\frac{1}{\pi500^2}=1.2732\times 10^{-6}, \{K=1, \lambda_1=11\lambda_0\}, \{K=2, \lambda_1=\lambda_0, \lambda_2=10\lambda_0, B_2=1\}, \{K=3, \lambda_1=\lambda_0, \lambda_2=4\lambda_0, \lambda_3=6\lambda_0, B_3=1\}$), and (c) secrecy constraints ($K=2, r_{a}=50,  \lambda_1=\frac{1}{\pi500^2}=1.2732\times 10^{-6}, \lambda_2=15\lambda_1, B_2=1$).}
	\label{Fig.fig7}
	\vspace{-0.5cm}
\end{figure*}

 	1) The ergodic secrecy rate in \cite{HetnetSECLKMAS2016} for secure HetNet is compared with our results versus density of
 	eavesdroppers ($\lambda_e$). In addition to the difference between our scheme and \cite{HetnetSECLKMAS2016} in the use of NOMA,  the work in
 	\cite{HetnetSECLKMAS2016} differs in some other aspects. Contrary to the considered model, the first tier in
 	\cite{HetnetSECLKMAS2016} is equipped with massive MIMO, intra-tier interference is not taken in to account in the second tier, and the adoption of S-FFR for inter-tier interference mitigation is regarded. It is assumed in S-FFR that there are a total of $K$ resource blocks; $\alpha K$ of them are allocated to small BSs and $(1-\alpha)K$ are shared by small BSs and macro BSs, where $\alpha$ denotes the S-FFR factor. In Fig. \ref{Fig.fig7}(a), the lower bound on ergodic secrecy rate of \cite{HetnetSECLKMAS2016} is presented by eliminating massive MIMO in the first tier and setting $\alpha=0$ in S-FFR. The results of our considered scheme with and without NOMA are denoted as SN-Het and HetNet with secrecy (by setting $a_m=0,r_s=0$), respectively. Note that secure HetNet, which is lower bound, is lower than HetNet with secrecy.  In accordance to Fig. \ref{Fig.fig7}(a), the SN-Het has superiority over other methods due to enabaled NOMA technique. 
 	%Since intra-tier interference is not considered in derivation of \cite{HetnetSECLKMAS2016}, the ergodic secrecy rate in \cite{HetnetSECLKMAS2016} is higher than HetNet with secrecy.

 	2) Secure multi-cell NOMA is prepared by eliminating HetNet from the SN-Het ($K=1$) to study the effect of using multi-tiers with a fixed density of BSs  in Fig. \ref{Fig.fig7}(b). It is observed that the ergodic secrecy rate in three-tier HetNet ($K=3$) has a significant performance improvement compared to multi-cell NOMA ($K=1$). This is consistent with the result of \cite{HetnetMAG2013} which does not include NOMA and secrecy.

	3) The ergodic rate of HetNets from \cite{HetnetHYPJ2012} and the lower bound on the ergodic rate of  NOMA-HetNet  from \cite{HetnetNOMAYZMAJ2017} are provided in Fig. \ref{Fig.fig7}(c) to study the effect of secrecy constraints. Massive MIMO in the first tier of model in \cite{HetnetNOMAYZMAJ2017} is eliminated to be comparable with our model. Moreover, we consider our scheme without NOMA (denoted as HetNet with secrecy) and eliminating NOMA only at the first tier (as the network in \cite{HetnetNOMAYZMAJ2017}, denoted as NOMA-HetNet with secrecy). In both with and without NOMA cases, secrecy constraint degrades the ergodic rates, especially at higher $\lambda_e$s.  It is worth noting that NOMA-HetNet in $\lambda_e=0$ is  lower bound and lower than NOMA-HetNet with secrecy.

\section{Conclusions}
The physical layer security in a $K$-tier multi-cell HetNet with NOMA in all tiers is investigated.  BSs, legitimate users, and eavesdroppers are randomly distributed in the considered system model. First, analytical expressions for the ergodic secrecy rate over all tiers of the network are derived. Next a lower bound on the ergodic secrecy rate is obtained. In addition, closed-form expressions for the lower bounds on the ergodic rates of legitimate users are provided in a special case. Furthermore, the ergodic secrecy rate of the network is studied in the interference-limited regime. Experimental results demonstrate that NOMA compared to OMA improves the spectrum efficiency performance. It is also inferred that  applying HetNet to a secure multi-cell NOMA system improves the spectrum efficiency performance. Moreover, the obtained results verify that increase in the number of network tiers with a fixed  density of BSs improves the ergodic secrecy rate. The ergodic secrecy rate is also enhanced with increase in density of the second tier BSs in a two-tier network while density of BSs at the first tier is kept fix. However, the secrecy constraint degrades the ergodic rate as expected.
 
\appendix
\numberwithin{equation}{subsection}
 \setcounter{equation}{0}  % reset counter 
\subsection{Proof of Lemma \ref{le1}} \label{Th2:Proof} %%\ref{newapp}
  The ergodic rate of NOMA users in the $k$-th tier is computed as follows:
	\begin{align}\label{eq:A.01}
	&{R_{w,k}}=\E_{(r_s,r_a)}\E_{\gamma_{{w}_{k,q}}}[\Theta(\gamma_{{w}_{k,q}}(r_s,r_a))]\nonumber\\
	&=\int_{0}^{\infty}\int_{0}^{\infty}\E_{\gamma_{{w}_{k,q}}}[\Theta(\gamma_{{w}_{k,q}}(r_s,r_a))]
	\times{}{f_{R_{s},R_{a}}(r_s,r_a)}\,dr_s\,dr_a\nonumber\\
	&=\int_{0}^{\infty}\int_{0}^{r_a}\E_{\gamma_{{w}_{k,q}}^{\text{\RNum{1}}}}[\Theta(\gamma_{{w}_{k,q}}^{\text{\RNum{1}}}(r_s,r_a))]\times{}{f_{R_{s}}(r_s|r_a)}{f_{R_{a}}(r_a)}\,dr_s\,dr_a\nonumber\\
	&+\int_{0}^{\infty}\int_{r_a}^{\infty}\E_{\gamma_{{w}_{k,q}}^{\text{\RNum{2}}}}[\Theta(\gamma_{{w}_{k,q}}^{\text{\RNum{2}}}(r_s,r_a))]\times{}{f_{R_{s}}(r_s|r_a)}{f_{R_{a}}(r_a)}\,dr_s\,dr_a\nonumber\\
	&\stackrel{(a)}{=}\int_{0}^{\infty}\int_{0}^{r_a}\big(\E_{({\gamma_{{m,m}_{k,q}}^{\text{\RNum{1}}},\gamma_{{n,m}_{k,q}}^{\text{\RNum{1}}}})}[\Theta(\text{min}(\gamma_{{m,m}_{k,q}}^{\text{\RNum{1}}}(r_s,r_a),\gamma_{{n,m}_{k,q}}^{\text{\RNum{1}}}\nonumber\\
	&(r_s,r_a)))]+\E_{\gamma_{{n,n}_{k,q}}^{\text{\RNum{1}}}}[\Theta(\gamma_{{n,n}_{k,q}}^{\text{\RNum{1}}}(r_s,r_a))]\big)
	\times{}{f_{R_{s}}(r_s)}{f_{R_{a}}(r_a)}\,dr_s\,dr_a\nonumber\\
	&+\int_{0}^{\infty}\int_{r_a}^{\infty}\big(\E_{({\gamma_{{m,m}_{k,q}}^{\text{\RNum{2}}},\gamma_{{n,m}_{k,q}}^{\text{\RNum{2}}}})}[\Theta(\text{min}(\gamma_{{m,m}_{k,q}}^{\text{\RNum{2}}}(r_s,r_a),\gamma_{{n,m}_{k,q}}^{\text{\RNum{2}}}\nonumber\\
	&(r_s,r_a)))]+\E_{\gamma_{{n,n}_{k,q}}^{\text{\RNum{2}}}}[\Theta(\gamma_{{n,n}_{k,q}}^{\text{\RNum{2}}}(r_s,r_a))]\big)\times{}{f_{R_{s}}(r_s)}{f_{R_{a}}(r_a)}\,dr_s\,dr_a\nonumber\\
	&={R_{m,k}^{\text{\RNum{1}}}}+{R_{n,k}^{\text{\RNum{1}}}}+{R_{m,k}^{\text{\RNum{2}}}}+{R_{n,k}^{\text{\RNum{2}}}},
	\end{align}
	where $\Theta(x)=log(1+x)$ and (a) caused by random selection of second user from network and $w\in\{m,n\}$.
	% Note that the upper limit of the integral on $r_s$ becomes $r_a$ in Case \RNum{1}. Also, the lower limit of the integral on  $r_s$ becomes $r_a$ in Case \RNum{2}.

 Keeping (\ref{eq:A.01}) in mind, the ergodic rate of the associated user ($m$) in Case \RNum{1} is calculated as:
\begin{align}\label{eq:A.02}
&{R_{m,k}^{\text{\RNum{1}}}}=\nonumber\\
&\int_{0}^{\infty}\int_{0}^{r_{a}}\E_{({\gamma_{{m,m}_{k,q}}^{\text{\RNum{1}}},\gamma_{{n,m}_{k,q}}^{\text{\RNum{1}}}})}[\Theta(\text{min}(\gamma_{{m,m}_{k,q}}^{\text{\RNum{1}}}(r_s,r_a),\gamma_{{n,m}_{k,q}}^{\text{\RNum{1}}}\nonumber\\
&(r_s,r_a)))]\times{}{f_{R_{s}}(r_s)}{f_{R_{a}}(r_a)}\,dr_s\,dr_a=\nonumber\\
&\int_{0}^{\infty}\int_{0}^{r_{a}}\Big\{\int_{0}^{\infty}\Pr\Big(\Theta(\text{min}(\gamma_{{m,m}_{k,q}}^{\text{\RNum{1}}}(r_s,r_a),\gamma_{{n,m}_{k,q}}^{\text{\RNum{1}}}(r_s,r_a)))\nonumber\\
&>y\Big)\,dy\Big\}\times{}{f_{R_{s}}(r_s)}{f_{R_{a}}(r_a)}\,dr_s\,dr_a\stackrel{(a)}{=}\nonumber\\
&\int_{0}^{\infty}\int_{0}^{r_{a}}(\frac{1}{\ln2}\int_{0}^{\infty}\frac{\bar {F} _{\text{min}(\gamma_{{m,m}_{k,q}}^{\text{\RNum{1}}},\gamma_{{n,m}_{k,q}}^{\text{\RNum{1}}})}(t)}{1+t}\,dt)\nonumber\\
&\times{}{f_{R_{s}}(r_s)}{f_{R_{a}}(r_a)}\,dr_s\,dr_a=\nonumber\\
&\frac{1}{\ln2}\int_{0}^{\infty}\int_{0}^{r_{a}}\int_{0}^{\infty}\frac{\Pr\Big({\text{min}(\gamma_{{m,m}_{k,q}}^{\text{\RNum{1}}},\gamma_{{n,m}_{k,q}}^{\text{\RNum{1}}})}>t\Big)}{1+t}\nonumber\\
&\times{}{f_{R_{s}}(r_s)}{f_{R_{a}}(r_a)}\,dt\,dr_s\,dr_a\stackrel{(b)}{=}\nonumber\\
&\frac{1}{\ln2}\int_{0}^{\infty}\int_{0}^{r_{a}}\int_{0}^{\infty}\frac{\Pr\Big(\gamma_{{m,m}_{k,q}}^{\text{\RNum{1}}}>t\Big)\Pr\Big(\gamma_{{n,m}_{k,q}}^{\text{\RNum{1}}}>t\Big)}{1+t}\nonumber\\
&\times{}{f_{R_{s}}(r_s)}{f_{R_{a}}(r_a)}\,dt\,dr_s\,dr_a,
\end{align}
where (a) follows from the variable change $y=\Theta(t)$. $\bar {F} _{{\text{min}(\gamma_{{m,m}_{k,q}}^{\text{\RNum{1}}},\gamma_{{n,m}_{k,q}}^{\text{\RNum{1}}})}}$ denotes the complete cumulative distribution function (CCDF) of ${\text{min}(\gamma_{{m,m}_{k,q}}^{\text{\RNum{1}}},\gamma_{{n,m}_{k,q}}^{\text{\RNum{1}}})}$ and  (b) results from independence of $\gamma_{{m,m}_{k,q}}^{\text{\RNum{1}}}$ and $\gamma_{{n,m}_{k,q}}^{\text{\RNum{1}}}$ because of a fixed and independent $r_a$ from $r_s$.  Defining $\gamma_{{m,m}_{k,q}}$ for Case~ \RNum{1} by substituting $\{{d_{m_{k,q}}}=r_{a}\}$ in (\ref{eq:05}), we have
\begin{align}\label{eq:A.03}
&\Pr\Big({\gamma_{{m,m}_{k,q}}^{\text{\RNum{1}}}}>t\Big)\nonumber\\
&=\Pr\Big({\frac{{a_{m_{k,q}}}P_k{g_{m_{k,q}}}r_{a}^{-\alpha_k}}{a_{n_{k,q}}P_k{g_{m_{k,q}}}r_{a}^{-\alpha_k}+{I_{m_{k,q}}}+{\sigma^2}}>t}\Big)\nonumber\\
&={\exp\Big\{-\frac{t{\sigma^2}{r_{a}}^{\alpha_k}}{({a_{m_{k,q}}}-t{a_{n_{k,q}}})P_k}\Big\}}{{L_{I_m}}\Big\{\frac{{t}{r_{a}}^{\alpha_k}}{({a_{m_{k,q}}}-t{a_{n_{k,q}}})P_k}\Big\}},
\end{align}
where  $L_{I_m}$ is obtained in (\ref{eq:09}). Defining $\gamma_{{n,m}_{k,q}}$ for Case \RNum{1} by substituting $\{{d_{n_{k,q}}}=r_{s}\}$ in (\ref{eq:06}), we similarly obtain
\begin{align}\label{eq:A.04}
&\Pr\Big({\gamma_{{n,m}_{k,q}}^{\text{\RNum{1}}}}>t\Big)\nonumber\\
&={\exp\Big\{-\frac{t{\sigma^2}{r_{s}}^{\alpha_k}}{({a_{m_{k,q}}}-t{a_{n_{k,q}}})P_k}\Big\}}{{L_{I_n}}\Big\{\frac{{t}{r_{s}}^{\alpha_k}}{({a_{m_{k,q}}}-t{a_{n_{k,q}}})P_k}\Big\}},
\end{align}
where  $L_{I_n}$ is obtained in (\ref{eq:09}). By considering $({a_{m_{k,q}}}-t{a_{n_{k,q}}})\ge 0$, which leads to $({\frac{a_{m_{k,q}}}{a_{n_{k,q}}}})\ge t$ based on unavailability of $\Pr(-g_{m_{k,q}}>\frac{t(I_{{k,q}}+\sigma^2)}{P_kr_{a}^{-\alpha_k}})$, the upper limit of the integral on $t$ in (\ref{eq:A.02}) becomes ${\frac{a_{m_{k,q}}}{a_{n_{k,q}}}}$. Hence, (\ref{eq:11}) is obtained by substituting (\ref{eq:A.03}) and (\ref{eq:A.04}) into (\ref{eq:A.02}).

Keeping (\ref{eq:A.01}) in mind, the ergodic rate of the second user ($n$) in Case \RNum{1} is similarly written as:
\begin{align}\label{eq:A.05}
{R_{n,k}^{\text{\RNum{1}}}}&=\int_{0}^{\infty}\int_{0}^{r_{a}}\E_{\gamma_{{n,n}_{k,q}}^{\text{\RNum{1}}}}[\Theta(\gamma_{{n,n}_{k,q}}^{\text{\RNum{1}}}(r_s,r_a))]\nonumber\\
&\times{}{f_{R_{s}}(r_s)}{f_{R_{a}}(r_a)}\,dr_s\,dr_a\nonumber\\
&=\int_{0}^{\infty}\int_{0}^{r_{a}}(\frac{1}{\ln2}\int_{0}^{\infty}\frac{\bar {F} _{\gamma_{{n,n}_{k,q}}^{\text{\RNum{1}}}}(t)}{1+t}\,dt)\nonumber\\
&\times{}{f_{R_{s}}(r_s)}{f_{R_{a}}(r_a)}\,dr_s\,dr_a\nonumber\\
&=\frac{1}{\ln2}\int_{0}^{\infty}\int_{0}^{r_{a}}\int_{0}^{\infty}\frac{\Pr\Big(\gamma_{{n,n}_{k,q}}^{\text{\RNum{1}}}>t\Big)}{1+t}\nonumber\\
&\times{}{f_{R_{s}}(r_s)}{f_{R_{a}}(r_a)}\,dt\,dr_s\,dr_a.
\end{align}
 Then, based on defining $\gamma_{{n,n}_{k,q}}$ for Case \RNum{1} by substituting $\{{d_{n_{k,q}}}=r_{s}\}$ in (\ref{eq:07}), we have
\begin{align}\label{eq:A.06}
\Pr\Big({\gamma_{{n,n}_{k,q}}^{\text{\RNum{1}}}}>t\Big)&={\exp\Big\{-\frac{t{\sigma^2}{r_{s}}^{\alpha_k}}{{a_{n_{k,q}}}P_k}\Big\}{{L_{I_n}}\Big\{\frac{{t}{r_{s}}^{\alpha_k}}{{a_{n_{k,q}}}P_k}\Big\}}}.
\end{align}
 By substituting (\ref{eq:A.06}) into  (\ref{eq:A.05}), we have
\begin{align}\label{A.07}
{R_{n,k}^{\text{\RNum{1}}}}&=\frac{1}{\ln2}\int_{0}^{\infty}\int_{0}^{r_{a}}\int_{0}^{\infty}{\frac{1}{1+t}}{\exp\Big\{-\frac{t{\sigma^2}{r_{s}}^{\alpha_k}}{{a_{n_{k,q}}}P_k}\Big\}}\nonumber\\
&\times{}{{L_{I_n}}\Big\{\frac{{t}{r_{s}}^{\alpha_k}}{{a_{n_{k,q}}}P_k}\Big\}}{f_{R_s}(r_{s})}{f_{R_a}(r_{a})}\,dt\,dr_s\,dr_a\nonumber\\
&=\frac{1}{\ln2}\int_{0}^{\infty}\int_{0}^{\infty}{\frac{1}{1+t}}{\exp\Big\{-\frac{t{\sigma^2}{r_{s}}^{\alpha_k}}{{a_{n_{k,q}}}P_k}\Big\}}{{L_{I_n}}\Big\{\frac{{t}{r_{s}}^{\alpha_k}}{{a_{n_{k,q}}}P_k}\Big\}}\nonumber\\
&\times{}{f_{R_s}(r_{s})}\,dt\,dr_s\int_{r_s}^{\infty}{(f_{R_a}(r_{a}))}\,dr_a\nonumber\\
&=\frac{1}{\ln2}\int_{0}^{\infty}\int_{0}^{\infty}{\frac{1}{1+t}}{\exp\Big\{-\frac{t{\sigma^2}{r_{s}}^{\alpha_k}}{{a_{n_{k,q}}}P_k}\Big\}}{{L_{I_n}}\Big\{\frac{{t}{r_{s}}^{\alpha_k}}{{a_{n_{k,q}}}P_k}\Big\}}\nonumber\\
&\times{}{(1-F_{R_a}(r_{s}))}{f_{R_s}(r_{s})}\,dt\,dr_s.
\end{align}
  Thus, (\ref{eq:12}) is obtained. The proof of (\ref{eq:13}) and (\ref{eq:14}) is similar to (\ref{eq:11}) and (\ref{eq:12}), respectively. This completes the proof.

 \setcounter{equation}{0} 
\subsection{Proof of Lemma \ref{le2}} 
The ergodic rate of legitimate user in Case \RNum{1} or Case \RNum{2} is ${{R}_{w,k}}=\E[\log(1+{(\gamma_{{w}_{k,q}})})]$ with $w\in\{m,n\}$, which is  expressed in (\ref{eq:11})-(\ref{eq:14}). The Jensen's inequality is utilized to derive a lower bound on the ergodic rate of the $w$-th user  as below: 
\begin{align}\label{eq:B.01}
{{R}_{w,k}}\ge\log(1+\frac{1}{{\E[\frac{1}{\gamma_{{w}_{k,q}}}]}}).
\end{align}
 Thus, the right hand side of (\ref{eq:B.01}) is the lower bound on the ergodic rate of legitimate user in Case \RNum{1} or Case \RNum{2} (i.e., ${\bar{R}_{w,k}}$), which is verified in this proof. Considering $\gamma_{{m,m}_{k,q}}$  and $\gamma_{{n,m}_{k,q}}$ for Case \RNum{1} obtained by substituting $\{{d_{m_{k,q}}}=r_{a}\}$ in (\ref{eq:05}) and $\{{d_{n_{k,q}}}=r_{s}\}$ in (\ref{eq:06}) as well as inequality $\frac{uv}{u+v}\le\min(u,v)$ that results in $u,v\ge0$ \cite{equalityAK2004}, the lower bound on the ergodic rate of the $m$-th user in Case~ \RNum{1} (${\bar{R}_{m,k}^{\text{\RNum{1}}}}$) is calculated at (\ref{eq:B.02}).

\begin{figure*}[!t]
			  \vspace{-0.4cm}
\begin{align}\label{eq:B.02}
	&\E\Big[\frac{1}{\text{min}(\gamma_{{m,m}_{k,q}}^{\text{\RNum{1}}},\gamma_{{n,m}_{k,q}}^{\text{\RNum{1}}})}\Big]\le\E\Big[\frac{\gamma_{{m,m}_{k,q}}^{\text{\RNum{1}}}+\gamma_{{n,m}_{k,q}}^{\text{\RNum{1}}}}{\gamma_{{m,m}_{k,q}}^{\text{\RNum{1}}}\gamma_{{n,m}_{k,q}}^{\text{\RNum{1}}}}\Big]\le\E\Big[{\frac{1}{\gamma_{{n,m}_{k,q}}^{\text{\RNum{1}}}}}+{\frac{1}{\gamma_{{m,m}_{k,q}}^{\text{\RNum{1}}}}}\Big]\nonumber\\
	&\le\E\Big[{\frac{a_{n_{k,q}}P_k{g_{n_{k,q}}}r_{s}^{-\alpha_k}+{I_{n_{k,q}}}+{\sigma^2}}{{a_{m_{k,q}}}P_k{g_{n_{k,q}}}r_{s}^{-\alpha_k}}}+{\frac{a_{n_{k,q}}P_k{g_{m_{k,q}}}r_{a}^{-\alpha_k}+{I_{m_{k,q}}}+{\sigma^2}}{{a_{m_{k,q}}}P_k{g_{m_{k,q}}}r_{a}^{-\alpha_k}}}\Big]\nonumber\\
	&\le\frac{2{a_{n_{k,q}}}}{{a_{m_{k,q}}}}+{\frac{1}{{{a_{m_{k,q}}}{P_k}}}}\Big(
	\int_{0}^{\infty}\int_{0}^{r_a}\E[\frac{1}{{g_{n_{k,q}}}}]\E[(I_{n_{k,q}}+{\sigma^2})]{{r_s}^{\alpha_k}}{f_{R_s}(r_s)}{f_{R_a}(r_{a})}\,dr_s\,dr_a
+\int_{0}^{\infty}\E[\frac{1}{{g_{m_{k,q}}}}]\E[(I_{m_{k,q}}+{\sigma^2})]{{r_a}^{\alpha_k}}{f_{R_a}(r_a)}\,dr_a\Big)\nonumber\\
	&\stackrel{(a)}{\le}\frac{2{a_{n_{k,q}}}}{{a_{m_{k,q}}}}+{\frac{1}{{{a_{m_{k,q}}}{P_k}}}}\Big(\int_{0}^{\infty}\big\{\E[{I_{n_{k,q}}}]+{\sigma^2}\big\}{{r_s}^{\alpha_k}}(1-{F_{R_a}(r_s)}){f_{R_s}(r_s)}\,dr_s+\int_{0}^{\infty}\big\{\E[{I_{m_{k,q}}}]+{\sigma^2}\big\}{{r_a}^{\alpha_k}}{f_{R_a}(r_a)}\,dr_a\Big),
\end{align}
\hrulefill
\end{figure*}

In (\ref{eq:B.02}), (a) results from $\E[\frac{1}{{g_{w_{k,q}}}}]\ge{\frac{1}{\E[{g_{w_{k,q}}}]}}$ due to the convexity of $\frac{1}{{g_{w_{k,q}}}}$ and considering ${g_{w_{k,q}}}\sim \exp(1)$. Based on (\ref{eq:05}) and (\ref{eq:06}), $\E[{I_{w_{k,q}}}]$ can be calculated as:
\begin{align}\label{eq:B.03}
\E[{I_{w_{k,q}}}]&=\E{\Big[{\sum_{j=1}^K \hspace{0.1cm}\sum_{l\in\Phi_{b_j}\backslash\left \{BS_{k,q} \right \}}{P_j{g_{w_{j,l}}}d_{w_{j,l}}^{-\alpha_j}}}\Big]}\nonumber\\
&\stackrel{(a)}{=}\sum_{j=1}^K(2\pi P_j{\lambda_{j}}\int_{y_{j}(r_u)}^{\infty}{z^{1-\alpha_j}}\,dz)\nonumber\\
&={\sum_{j=1}^K(\frac{2\pi P_j{\lambda_{j}}}{\alpha_j-2}){y_{j}(r_u)}^{2-\alpha_j}},
\end{align}
where (a)  results from using Campbell's theorem and ${g_{w_{j,l}}}\sim \exp(1)$. Using $\E[{I_{w_{k,q}}}]$ in (\ref{eq:B.03}), (\ref{eq:15}) is obtained by substituting (\ref{eq:B.02}) into the right hand side of (\ref{eq:B.01}).

Similarly, to calculate the lower bound on the ergodic rate of the $n$-th user in Case \RNum{1} (${\bar{R}_{n,k}^{\text{\RNum{1}}}}$), based on defining $\gamma_{{n,n}_{k,q}}$ for Case \RNum{1} by substituting $\{{d_{n_{k,q}}}=r_{s}\}$ in (\ref{eq:07}), we have:
\begin{align}\label{eq:B.04}
&\E[\frac{1}{\gamma^{\text{\RNum{1}}}_{{n,n}_{k,q}}}]=\E[\frac{({I_{n_{k,q}}}+{\sigma^2})r_{s}^{\alpha_k}}{{a_{n_{k,q}}}P_k{g_{n_{k,q}}}}]\stackrel{(a)}{=}\nonumber\\
&\frac{1}{{{a_{n_{k,q}}}{P_k}}}\int_{0}^{\infty}\big\{\E[{I_{n_{k,q}}}]+{\sigma^2}\big\}{{r_s}^{\alpha_k}}(1-{F_{R_a}(r_s)}){f_{R_s}(r_s)}\,dr_s,
\end{align}
 where (a) is derived in (\ref{eq:B.02}). Again, using $\E[{I_{w_{k,q}}}]$ of (\ref{eq:B.03}), (\ref{eq:16}) is achieved by substituting (\ref{eq:B.04}) into the right hand side of (\ref{eq:B.01}). The proof procedure for the lower bounds on the ergodic rates of the $m$-th and $n$-th users in Case \RNum{2} is similar to  Case \RNum{1}. This completes the proof.

\setcounter{equation}{0} 
\subsection{Proof of Corollary \ref{co1}} 
Substituting $\alpha_k=\alpha,k\in[1:K]$ and  $\hat{B_j}=1$ in (\ref{eq:15}), the lower bound on the ergodic rate of the $m$-th user in Case \RNum{1} is:
\begin{align}\label{eq:C.01}
{\tilde{R}_{m,k}^{\text{\RNum{1}}}}&=\log\Big(1+\Big(\frac{2{a_{n_{k,q}}}}{{a_{m_{k,q}}}}+{\frac{1}{{{a_{m_{k,q}}}{P_k}}}}\nonumber\\
&\times{}\underbrace{\int_{0}^{\infty}\big\{\E[{I_{n_{k,q}}}]+{\sigma^2}\big\}{{r_s}^{\alpha}}(1-{F_{R_a}(r_s)}){f_{R_s}(r_{s})}\,dr_s}_{{C}}\nonumber\\
&+{\frac{1}{{{a_{m_{k,q}}}{P_k}}}}\underbrace{\int_{0}^{\infty}\big\{\E[{I_{m_{k,q}}}]+{\sigma^2}\big\}{{r_a}^{\alpha}}{f_{R_a}(r_a)}\,dr_a}_{{D}}\Big)^{-1}\Big).
\end{align}
First, $\E[I_{w_{k,q}}]$ from (\ref{eq:B.03}) is rewritten as:
\begin{align}\label{eq:C.02}
\E[{I_{w_{k,q}}}]\stackrel{(a)}{=}\frac{2 P_k}{\alpha-2}{{E}}{r_u}^{2-\alpha},
\end{align}
where (a) results from $y_{j}(r_u)={\hat{P_j}^{1/\alpha}}{r_u}$ and  ${E}\triangleq\sum_{j=1}^{K}{\pi\lambda_{j}}\hat{P_j}^{\frac{2}{\alpha}}$.

\noindent
 Based on $\alpha_k=\alpha,k\in[1:K]$ and  $\hat{B_j}=1$,  $A_k$ in (\ref{eq:03}) is calculated as:
\begin{align}\label{eq:C.03}
{A_{k}}&={2\pi\lambda_{k}}\int_{0}^{\infty} {r}\exp\{-{E}{{r}^{2}}\}\,dr=\frac{\lambda_{k}}{\sum_{j=1}^{K}{\lambda_{j}}\hat{P_j}^{\frac{2}{\alpha}}},
\end{align}
Then, we have 
\begin{align}\label{eq:C.04}
\frac{\lambda_{k}}{A_k}=\sum_{j=1}^{K}{\lambda_{j}}\hat{P_j}^{\frac{2}{\alpha}}.
\end{align} 
Considering $w=n$ in (\ref{eq:C.02}),  $u=s$ in (\ref{eq:04}), and using (\ref{eq:C.04}), $C$ can be calculated as follows:
\begin{align}\label{eq:C.05}
C&=\int_{0}^{\infty}{{2E}}({\frac{2P_k}{\alpha-2}}{E}{{r_s}^{2-\alpha}}+{\sigma^2}){{r_s}^{\alpha+1}}\exp\{-{E}{{r_s}^2}\}\nonumber\\
&\times{}(1-{F_{R_a}(r_s)})\,dr_s\nonumber\\
&=\int_{0}^{\infty}{\frac{4P_k{{E}^2}}{\alpha-2}}{{r_s}^{3}}\exp\{-{2E}{{r_s}^2}\}\,d{r_s}\nonumber\\
&+\int_{0}^{\infty}{{2E}}{\sigma^2}{{r_s}^{\alpha+1}}\exp\{-{2E}{{r_s}^2}\}\,d{r_s}\nonumber\\
&\stackrel{(a)}{=}{\frac{P_k}{2(\alpha-2)}}+{\frac{{\sigma^2}\Gamma({\frac{\alpha}{2}+1})}{2(2{E})^{\frac{\alpha}{2}}}},
\end{align}
where utilizing \cite[Eq. (3.326.2)]{BookII2007} leads to (a). Then, considering $w=m$ in (\ref{eq:C.02}), $u=a$ in (\ref{eq:04}), and exploiting (\ref{eq:C.04}), $D$ can be calculated as below:
\begin{align}\label{eq:C.06}
D&=\int_{0}^{\infty}\Big({\frac{4P_k{E^2}}{\alpha-2}}{r_a^{3}}\exp\{-{E}{r_a^2}\}
+{2{E}\sigma^2}{r_a^{\alpha+1}}\exp\{-{E}{r_a^2}\}\Big)\,dr_a\nonumber\\
&\stackrel{(a)}{=}\frac{2P_k}{\alpha-2}+\frac{{\sigma^2}\Gamma({\frac{\alpha}{2}}+1)}{{{E}^{\frac{\alpha}{2}}}},
\end{align}
where (a) is defined in (\ref{eq:C.05}). Eventually, (\ref{eq:19}) is achieved by substituting (\ref{eq:C.05}) and (\ref{eq:C.06}) into (\ref{eq:C.01}).

Similarly, by substituting $\alpha_k=\alpha,k\in[1:K]$ and  $\hat{B_j}=1$ in (\ref{eq:16}), the lower bound on the ergodic rate of the $n$-th user in Case \RNum{1} is written as:
\begin{align}\label{eq:C.07}
{\bar{R}_{n,k}^{\text{\RNum{1}}}}&=\log\Big(1+\Big(\frac{1}{{{a_{n_{k,q}}}{P_k}}}\times{}\nonumber\\
&\underbrace{\int_{0}^{\infty}\big\{\E[{I_{n_{k,q}}}]+{\sigma^2}\big\}{{r_s}^{\alpha}}(1-{F_{R_a}(r_s)}){f_{R_s}(r_s)}\,dr_s}_{{C}}\Big)^{-1}\Big),
\end{align}
where C is derived at (\ref{eq:C.05}). Finally, (\ref{eq:20}) is obtained by substituting (\ref{eq:C.05}) into (\ref{eq:C.07}). The proof procedure for (\ref{eq:21}) and (\ref{eq:22}) is similar to (\ref{eq:19}) and (\ref{eq:20}), respectively. This completes the proof.

 \setcounter{equation}{0} 
\subsection{Proof of Lemma \ref{le3}} 
The ergodic leakage rate of the most detrimental eavesdropper for detecting information of the $w$-th user ($w\in\{m,n\}$) is written as:
\begin{align}\label{eq:D.01}
\E[\log(1+\gamma_{e_{max_{k,q}}})]&=\frac{1}{\ln2}\int_{0}^{\infty}\frac{1-{F} _{\gamma_{e_{max_{k,q}}}}(t)}{1+t}\,dt.
\end{align}
Based on (\ref{eq:08}), the CDF of $\gamma_{e_{max_{k,q}}}$ is calculated as:
\begin{align}\label{eq:D.02}
&{F}_{\gamma_{e_{max_{k,q}}}}(t)=\Pr\Big(\gamma_{e_{max_{k,q}}}<t\Big)=\nonumber\\
&\Pr\Big(\max_{e\in\Phi_{e}}\frac{{a_{w_{k,q}}}P_k{g_{e_{k,q}}}d_{e_{k,q}}^{-\alpha_k}}{{I_{e_{k,q}}}+{\sigma^2}}<t\Big)=\nonumber\\
&\E_{\Phi_{e}}\Bigg[\prod_{e\in\Phi_{e}}\Pr\Big(\frac{{a_{w_{k,q}}}P_k{g_{e_{k,q}}}d_{e_{k,q}}^{-\alpha_k}}{{I_{e_{k,q}}}+{\sigma^2}}<t \mid \Phi_{e}\Big)\Bigg]\stackrel{(a)}{=}\nonumber\\
&\exp\Bigg\{-\lambda_e\int_{0}^{\infty}1-\Pr\Big(\frac{{a_{w_{k,q}}}P_k{g_{e_{k,q}}}r^{-\alpha_k}}{{I_{e_{k,q}}}+{\sigma^2}}<t\Big)\,dr\Bigg\}\stackrel{(b)}{=}\nonumber\\
&\exp\Bigg\{-2\pi\lambda_e\int_{0}^{\infty}\Pr\Big({g_{e_{k,q}}}>{\frac{t({I_{e_{k,q}}}+{\sigma^2})}{{a_{w_{k,q}}}P_kr^{-\alpha_k}}}\Big)r\,dr\Bigg\}=\nonumber\\
&\exp\Bigg\{-2\pi\lambda_e\int_{0}^{\infty}{\exp\Big\{-\frac{t{\sigma^2}{r}^{\alpha_k}}{{a_{w_{k,q}}}P_k}\Big\}}{{L_{I_e}}\Big\{\frac{{t}r^{\alpha_k}}{{a_{w_{k,q}}}P_k}\Big\}}r\,dr\Bigg\},
\end{align}
where (a) and (b) are in accordance to use of the generating functional of the PPP $\Phi_{e}$ and the polar-coordinate system, respectively;  $L_{I_e}$ is obtained by (\ref{eq:10}). Finally, (\ref{eq:23}) is obtained by substituting (\ref{eq:D.02}) into (\ref{eq:D.01}). This completes the proof.

\setcounter{equation}{0} 
\subsection{Proof of Corollary \ref{co2}} 
Substituting $\sigma^2=0$, $\alpha_k=\alpha,k\in[1:K]$, and  $\hat{B_j}=1$ in (\ref{eq:11}), the ergodic rate of the associated user ($m$) in Case \RNum{1} is:
\begin{align}\label{eq:E.01}
&{R_{m,k}^{\text{\RNum{1}}}(\alpha,1)}\nonumber\\
&=\frac{1}{\ln2}\int_{0}^{\infty}\int_{0}^{r_{a}}\int_{0}^{\frac{a_{m_{k,q}}}{a_{n_{k,q}}}}{\frac{1}{1+t}}{{L_{I_m}}\Big\{\frac{{t}{r_{a}}^{\alpha}}{({a_{m_{k,q}}}-t{a_{n_{k,q}}})P_k}\Big\}}\nonumber\\
&\times{}{{L_{I_n}}\Big\{\frac{{t}{r_{s}}^{\alpha}}{({a_{m_{k,q}}}-t{a_{n_{k,q}}})P_k}\Big\}}{f_{R_s}(r_{s})}{f_{R_a}(r_{a})}\,dt\,dr_s\,dr_a,
\end{align}
where from (\ref{eq:09}) we obtain:

\begin{align}\label{eq:E.02}
&{L_{I_m}}_{(\alpha,1)}{\Big\{\frac{{t}{r_{a}}^{\alpha}}{\left({a_{m_{k,q}}}-t{a_{n_{k,q}}}\right)P_k}\Big\}}\stackrel{(a)}{=}\exp\Bigg\{-\sum_{j=1}^{K}\pi{\lambda_{j}}{r_{a}^2}\hat{P_j}^{\frac{2}{\alpha}}\nonumber\\
&\frac{2tA}{\alpha-2}\times{}_2\!F_1\Big(1,1-\frac{2}{\alpha};2-\frac{2}{\alpha};-{tA}\Big)\Bigg\}\stackrel{(b)}{=}\nonumber\\
&\exp\Bigg\{-\sum_{j=1}^{K}\pi{\lambda_{j}}{r_{a}^2}\hat{P_j}^{\frac{2}{\alpha}}Z_m\Bigg\},
\end{align}
  where (a) and (b) follow from $A\triangleq\frac{1}{{a_{m_{k,q}}}-t{a_{n_{k,q}}}}$ and $Z_m\triangleq \frac{2tA}{\alpha-2}\times{}_2\!F_1(1,1-\frac{2}{\alpha};2-\frac{2}{\alpha};-{tA})$, respectively. Similarly, we use (\ref{eq:09}) to obtain:
 \begin{align}\label{eq:E.03}
 {L_{I_n}}_{(\alpha,1)}{\Big\{\frac{{t}{r_{s}}^{\alpha}}{\left({a_{m_{k,q}}}-t{a_{n_{k,q}}}\right)P_k}\Big\}}=\exp\Bigg\{-\sum_{j=1}^{K}\pi{\lambda_{j}}{r_{s}^2}\hat{P_j}^{\frac{2}{\alpha}}Z_m\Bigg\}.
 \end{align} 
% Based on $\sigma^2=0$, $\left\{\alpha_k\right\}_{k=1,...,K}=\alpha$, and  $\hat{B_j}=1$,  $A_k$ in (\ref{eq:03}) is calculated as
% \begin{align}\label{eq:E.04}
 %{A_{k}}\stackrel{(a)}{=}{2\pi\lambda_{k}}\int_{0}^{\infty} {r_u}\exp\Big\{-{D}{{r_u}^{2}}\Big\}\,dr_u\stackrel{(b)}{=}{\pi\lambda_{k}}\int_{0}^{\infty}\exp\Big\{-{D}t\Big\}\,dt=\frac{\lambda_{k}}{\sum_{j=1}^{K}{\lambda_{j}}\hat{P_j}^{\frac{2}{\alpha}}},
 %\end{align}
 %where (a) follows from ${D}\triangleq\sum_{j=1}^{K}{\pi\lambda_{j}}\hat{P_j}^{\frac{2}{\alpha}}$ and (b) results from substitution of ${r_u}^2=t$. Then, we have 
 %\begin{align}\label{eq:E.05}
 %\frac{\lambda_{k}}{A_k}=\sum_{j=1}^{K}{\lambda_{j}}\hat{P_j}^{\frac{2}{\alpha}}.
 %\end{align}
 By substituting (\ref{eq:E.02}) and (\ref{eq:E.03}) into  (\ref{eq:E.01}), $u=s$ in (\ref{eq:04}), and exploiting (\ref{eq:C.04}), (\ref{eq:E.01}) can be rewritten as:
\begin{align}\label{eq:E.04}
&{R_{m,k}^{\text{\RNum{1}}}(\alpha,1)}=\frac{1}{\ln2}\int_{0}^{\infty}\int_{0}^{r_{a}}\int_{0}^{\frac{a_{m_{k,q}}}{a_{n_{k,q}}}}\frac{\sum_{j=1}^{K}{2\pi\lambda_{j}}\hat{P_j}^{\frac{2}{\alpha}}r_{s}}{1+t}\nonumber\\
&\times{}\underbrace{\exp\Bigg\{-{\sum_{j=1}^{K}{\pi\lambda_{j}}\hat{P_j}^{\frac{2}{\alpha}}}\Big({Z_m{r_a}^2}+{(1+Z_m){r_s}^2}\Big)\Bigg\}}_{{G}}\nonumber\\
&\times{}{f_{R_{a}}(r_a)}\,dt\,dr_s\,dr_a.
\end{align}
The inner integral can be calculated as:
 \begin{align}\label{eq:E.05}
  &\int_{0}^{r_{a}}r_{s}{G}\,dr_s\nonumber\\
  &\stackrel{(a)}{=}\int_{0}^{r_{a}}r_{s}\exp\Big\{-{E}(1+Z_m)r_{s}^2\Big\}\exp\Big\{-{E}{r_{a}}^{2}Z_m\Big\}\,dr_s\nonumber\\
  &=\frac{\exp\Big\{-{E}{r_{a}}^{2}Z_m\Big\}}{2{E}(1+Z_m)}\Big(1-\exp\Big\{{-{E}(1+Z_m){r_{a}}^{2}}\Big\}\Big),
  \end{align}
where (a) follows from  $E\triangleq\sum_{j=1}^{K}{\pi\lambda_{j}}\hat{P_j}^{\frac{2}{\alpha}}$ and ${G}$ is defined in (\ref{eq:E.04}). Thus, (\ref{eq:E.01}) can be rewritten by considering $u=a$ in (\ref{eq:04}) as follows:
  \begin{align}\label{eq:E.06}
 &{R_{m,k}^{\text{\RNum{1}}}(\alpha,1)}=\nonumber\\
 &\frac{1}{\ln2}\int_{0}^{\infty}\int_{0}^{\frac{a_{m_{k,q}}}{a_{n_{k,q}}}}\frac{2r_{a}{E}}{(1+t)(1+Z_m)}\exp\Big\{-{E}{r_{a}}^{2}Z_m\Big\}\nonumber\\
 &\times{}\Big(\exp\Big\{-{E}{r_{a}}^{2}\Big\}-\exp\Big\{-{E}{r_{a}}^{2}(2+Z_m)\Big\}\Big)\,dt\,dr_a.
  \end{align}
   Finally, (\ref{eq:24}) is achieved after some simplifications and utilizing  $\int_{0}^{\infty}r_{a}\exp\Big\{-{E}{r_{a}}^{2}\Big\}\,dr_a=\frac{1}{2{E}}$ based on \cite[Eq. (3.326.2)]{BookII2007}.

  Similarly, the ergodic rate of the second user ($n$) in Case \RNum{1} is written by substituting $\sigma^2=0$, $\alpha_k=\alpha,k\in[1:K]$, and  $\hat{B_j}=1$ in (\ref{eq:12}) as:
  \begin{align}\label{eq:E.07}
  {R_{n,k}^{\text{\RNum{1}}}(\alpha,1)}&=\frac{1}{\ln2}\int_{0}^{\infty}\int_{0}^{\infty}\frac{1}{1+t}\times{}\nonumber\\
  &{{L_{I_n}}_{(\alpha,1)}\Big\{\frac{{t}{r_{s}}^{\alpha}}{a_{n_{k,q}}P_k}\Big\}}(1-{F_{R_a}(r_{s})}){f_{R_s}(r_{s})}\,dt\,dr_s.
  \end{align}
   By substituting ${L_{I_n}}_{(\alpha,1)}{\Big\{\frac{{t}{r_{s}}^{\alpha}}{a_{n_{k,q}}P_k}\Big\}}$ from (\ref{eq:09})  into  (\ref{eq:E.07}), $u=a$ in (\ref{eq:04}), and using (\ref{eq:C.04}), (\ref{eq:E.07}) can be rewritten as:
  \begin{align}\label{eq:E.08}
  {R_{n,k}^{\text{\RNum{1}}}(\alpha,1)}&\stackrel{(a)}{=}\frac{1}{\ln2}\int_{0}^{\infty}\int_{0}^{\infty}\frac{\sum_{j=1}^{K}{2\pi\lambda_{j}}\hat{P_j}^{\frac{2}{\alpha}}r_{s}}{1+t}\times{}\nonumber\\
  &\underbrace{\exp\Bigg\{-{\sum_{j=1}^{K}{\pi\lambda_{j}}\hat{P_j}^{\frac{2}{\alpha}}}{r_{s}^2}\Big(2+Z_n\Big)\Bigg\}}_{{H}}{f_{R_{s}}(r_s)}\,dt\,dr_s,
  \end{align}
  where (a) follows from $Z_n\triangleq\frac{2t}{{a_{n_{k,q}}}(\alpha-2)}
  \times{}_2\!F_1(1,1-\frac{2}{\alpha};2-\frac{2}{\alpha};-\frac{t}{a_{n_{k,q}}})$.
  The inner integral can be computed by considering $u=s$ in (\ref{eq:04}) as:
  \begin{align}\label{eq:E.09} \int_{0}^{\infty}r_{s}{H}{f_{R_{s}}(r_s)}\,dr_s&=\int_{0}^{\infty}r_{s}\exp\Big\{-{E}(2+Z_n)r_{s}^2\Big\}\,dr_s\nonumber\\
  	&=\frac{1}{2{E}(2+Z_n)},
  \end{align}
   where ${H}$ is defined in (\ref{eq:E.08}). Finally, (\ref{eq:25}) is obtained based on \cite[Eq. (3.326.2)]{BookII2007} after doing some simplifications.

The ergodic leakage rate in the most detrimental eavesdropper for detecting information of the $w$-th user ($w\in\{m,n\}$) is obtained as follows by substituting $\sigma^2=0$, $\alpha_k=\alpha,k\in[1:K]$, and $\hat{B_j}=1$ in (\ref{eq:23}): 
\begin{align}\label{eq:E.10}
&{R_{e,k}^{w}(\alpha,1)}=\frac{1}{\ln2}\int_{0}^{\infty}\frac{1}{1+t}\times{}\nonumber\\
&\Big(1-\exp\Big\{-2\pi\lambda_e\int_{0}^{\infty}{L_{I_e}}_{(\alpha,1)}\Big\{\frac{{t}{d_{e_{k,q}}}^{\alpha}}{a_{w_{k,q}}P_k}\Big\}r\,dr\Big\}\Big)\,dt.
\end{align}
Substituting ${L_{I_e}}_{(\alpha,1)}{\Big\{\frac{{t}{d_{e_{k,q}}}^{\alpha}}{a_{n_{k,q}}P_k}\Big\}}$ from (\ref{eq:10}) into (\ref{eq:E.10}) gives: 
\begin{align}\label{eq:E.11}
&{R_{e,k}^{w}(\alpha,1)}\stackrel{(a)}{=}\frac{1}{\ln2}\int_{0}^{\infty}\frac{1}{1+t}\times{}\nonumber\\
&\Big(1-\exp\Big\{-2\pi\lambda_e\int_{0}^{\infty}\exp\Big\{-{D}_e{r^2}\Big\}r\,dr\Big\}\Big)\,dt,
\end{align}
where (a) results from defining ${D}_e\triangleq\sum_{j=1}^{K}\pi{\lambda_{j}}\hat{P_j}^{\frac{2}{\alpha}}(\frac{t}{a_{w_{k,q}}})^{\frac{2}{\alpha}}\Gamma(1+\frac{2}{\alpha})\Gamma(1-\frac{2}{\alpha})$. Using \cite[Eq. (3.326.2)]{BookII2007}, (\ref{eq:28}) is achieved after some simplifications. This completes the proof.

%%******-------------------------------------
%%******-------------------------------------
%\setLTRbibitems
%\resetlatinfont

\bibliographystyle{IEEEtranN}
\bibliography{FINAL}

% Generated by IEEEtranN.bst, version: 1.14 (2015/08/26)
\begin{thebibliography}{33}
\providecommand{\natexlab}[1]{#1}
\providecommand{\url}[1]{#1}
\csname url@samestyle\endcsname
\providecommand{\newblock}{\relax}
\providecommand{\bibinfo}[2]{#2}
\providecommand{\BIBentrySTDinterwordspacing}{\spaceskip=0pt\relax}
\providecommand{\BIBentryALTinterwordstretchfactor}{4}
\providecommand{\BIBentryALTinterwordspacing}{\spaceskip=\fontdimen2\font plus
\BIBentryALTinterwordstretchfactor\fontdimen3\font minus
  \fontdimen4\font\relax}
\providecommand{\BIBforeignlanguage}[2]{{%
\expandafter\ifx\csname l@#1\endcsname\relax
\typeout{** WARNING: IEEEtranN.bst: No hyphenation pattern has been}%
\typeout{** loaded for the language `#1'. Using the pattern for}%
\typeout{** the default language instead.}%
\else
\language=\csname l@#1\endcsname
\fi
#2}}
\providecommand{\BIBdecl}{\relax}
\BIBdecl

\bibitem[Alliance(2015)]{5GN2015}
N.~Alliance, ``{5G} white paper,'' \emph{Next generation mobile networks, white
  paper}, vol.~1, 2015.

\bibitem[Gupta and Jha(2015)]{5GAR2015}
A.~Gupta and R.~K. Jha, ``A survey of {5G} network: Architecture and emerging
  technologies,'' \emph{IEEE access}, vol.~3, pp. 1206--1232, 2015.

\bibitem[Agiwal et~al.(2016)Agiwal, Roy, and Saxena]{5GMAN2016}
M.~Agiwal, A.~Roy, and N.~Saxena, ``Next generation {5G} wireless networks: A
  comprehensive survey,'' \emph{IEEE Communications Surveys \& Tutorials},
  vol.~18, no.~3, pp. 1617--1655, 2016.

\bibitem[Brueck(2011)]{HetNetANJV2010}
S.~Brueck, ``Heterogeneous networks in {LTE}-advanced,'' in \emph{8th
  International Symposium on Wireless Communication Systems}, Aachen, Germany,
  2011, pp. 171--175.

\bibitem[Damnjanovic et~al.(2011)Damnjanovic, Montojo, Wei, Ji, Luo, Vajapeyam,
  Yoo, Song, and Malladi]{HetNetAJYTTMTOD2011}
A.~Damnjanovic, J.~Montojo, Y.~Wei, T.~Ji, T.~Luo, M.~Vajapeyam, T.~Yoo,
  O.~Song, and D.~Malladi, ``A survey on {3GPP} heterogeneous networks,''
  \emph{IEEE Wireless communications}, vol.~18, no.~3, pp. 10--21, 2011.

\bibitem[Ding et~al.(2017)Ding, Lei, Karagiannidis, Schober, Yuan, and
  Bhargava]{NOMAZXGRJV2017}
Z.~Ding, X.~Lei, G.~K. Karagiannidis, R.~Schober, J.~Yuan, and V.~K. Bhargava,
  ``A survey on non-orthogonal multiple access for {5G} networks: Research
  challenges and future trends,'' \emph{IEEE Journal on Selected Areas in
  Communications}, vol.~35, no.~10, pp. 2181--2195, 2017.

\bibitem[Dai et~al.(2018)Dai, Wang, Ding, Wang, Chen, and
  Hanzo]{NOMALBZZSL2018}
L.~Dai, B.~Wang, Z.~Ding, Z.~Wang, S.~Chen, and L.~Hanzo, ``A survey of
  non-orthogonal multiple access for {5G},'' \emph{IEEE communications surveys
  \& tutorials}, vol.~20, no.~3, pp. 2294--2323, 2018.

\bibitem[Yang et~al.(2015)Yang, Wang, Geraci, Elkashlan, Yuan, and
  Di~Renzo]{PLSNLGMJM2015}
N.~Yang, L.~Wang, G.~Geraci, M.~Elkashlan, J.~Yuan, and M.~Di~Renzo,
  ``Safeguarding {5G} wireless communication networks using physical layer
  security,'' \emph{IEEE Communications Magazine}, vol.~53, no.~4, pp. 20--27,
  2015.

\bibitem[Wu et~al.(2018)Wu, Khisti, Xiao, Caire, Wong, and Gao]{PLSYACGKX2018}
Y.~Wu, A.~Khisti, C.~Xiao, G.~Caire, K.-K. Wong, and X.~Gao, ``A survey of
  physical layer security techniques for {5G} wireless networks and challenges
  ahead,'' \emph{IEEE Journal on Selected Areas in Communications}, vol.~36,
  no.~4, pp. 679--695, 2018.

\bibitem[Liu et~al.(2016)Liu, Chen, and Wang]{PLSYHL2016}
Y.~Liu, H.-H. Chen, and L.~Wang, ``Physical layer security for next generation
  wireless networks: Theories, technologies, and challenges,'' \emph{IEEE
  Communications Surveys \& Tutorials}, vol.~19, no.~1, pp. 347--376, 2016.

\bibitem[Duong et~al.(2017)Duong, Zhou, and Poor]{PLSTXH2017}
T.~Q. Duong, X.~Zhou, and H.~V. Poor, \emph{\BIBforeignlanguage{English
  (US)}{Trusted communications with physical layer security for {5G} and
  beyond}}.\hskip 1em plus 0.5em minus 0.4em\relax United Kingdom: Institution
  of Engineering \& Technology, Jan 2017.

\bibitem[Zhang et~al.(2016)Zhang, Wang, Yang, and Ding]{NOMASECYHQZ2016}
Y.~Zhang, H.-M. Wang, Q.~Yang, and Z.~Ding, ``Secrecy sum rate maximization in
  non-orthogonal multiple access,'' \emph{IEEE Communications Letters},
  vol.~20, no.~5, pp. 930--933, 2016.

\bibitem[He et~al.(2017)He, Liu, Yang, and Lau]{NOMASECBANV2017}
B.~He, A.~Liu, N.~Yang, and V.~K. Lau, ``On the design of secure non-orthogonal
  multiple access systems,'' \emph{IEEE Journal on Selected Areas in
  Communications}, vol.~35, no.~10, pp. 2196--2206, 2017.

\bibitem[Liu et~al.(2017{\natexlab{a}})Liu, Qin, Elkashlan, Gao, and
  Hanzo]{NOMASECYZMYL2017}
Y.~Liu, Z.~Qin, M.~Elkashlan, Y.~Gao, and L.~Hanzo, ``Enhancing the physical
  layer security of non-orthogonal multiple access in large-scale networks,''
  \emph{IEEE Transactions on Wireless Communications}, vol.~16, no.~3, pp.
  1656--1672, 2017.

\bibitem[Chen et~al.(2018)Chen, Yang, and Alouini]{NOMASECJLM2018}
J.~Chen, L.~Yang, and M.-S. Alouini, ``Physical layer security for cooperative
  {NOMA} systems,'' \emph{IEEE Transactions on Vehicular Technology}, vol.~67,
  no.~5, pp. 4645--4649, 2018.

\bibitem[Zheng et~al.(2018)Zheng, Wen, Wang, Wang, Chen, Tang, and
  Ji]{NOMASECBMCXFJF2018}
B.~Zheng, M.~Wen, C.-X. Wang, X.~Wang, F.~Chen, J.~Tang, and F.~Ji, ``Secure
  {NOMA} based two-way relay networks using artificial noise and full duplex,''
  \emph{IEEE Journal on Selected Areas in Communications}, vol.~36, no.~7, pp.
  1426--1440, 2018.

\bibitem[Wang et~al.(2016{\natexlab{a}})Wang, Zheng, Yuan, Towsley, and
  Lee]{HetnetSECHTJDM2016}
H.-M. Wang, T.-X. Zheng, J.~Yuan, D.~Towsley, and M.~H. Lee, ``Physical layer
  security in heterogeneous cellular networks,'' \emph{IEEE Transactions on
  Communications}, vol.~64, no.~3, pp. 1204--1219, 2016.

\bibitem[Xu et~al.(2016)Xu, Tao, Yang, and Wu]{HetnetSECMXFH2016}
M.~Xu, X.~Tao, F.~Yang, and H.~Wu, ``Enhancing secured coverage with {CoMP}
  transmission in heterogeneous cellular networks,'' \emph{IEEE Communications
  Letters}, vol.~20, no.~11, pp. 2272--2275, 2016.

\bibitem[Wang et~al.(2016{\natexlab{b}})Wang, Wong, Elkashlan, Nallanathan, and
  Lambotharan]{HetnetSECLKMAS2016}
L.~Wang, K.-K. Wong, M.~Elkashlan, A.~Nallanathan, and S.~Lambotharan,
  ``Secrecy and energy efficiency in massive {MIMO} aided heterogeneous
  {C-RAN}: A new look at interference,'' \emph{IEEE Journal of Selected Topics
  in Signal Processing}, vol.~10, no.~8, pp. 1375--1389, 2016.

\bibitem[Wang et~al.(2017)Wang, Teh, and Li]{HetnetSECWKK2017}
W.~Wang, K.~C. Teh, and K.~H. Li, ``Artificial noise aided physical layer
  security in multi-antenna small-cell networks,'' \emph{IEEE Transactions on
  Information Forensics and Security}, vol.~12, no.~6, pp. 1470--1482, 2017.

\bibitem[Wang et~al.(2018)Wang, Gao, Dong, Sha, and Zang]{HetnetSECSYCNG2018}
S.~Wang, Y.~Gao, C.~Dong, N.~Sha, and G.~Zang, ``Secure user association in
  two-tier heterogeneous cellular networks with in-band interference,''
  \emph{IEEE Access}, vol.~6, pp. 38\,607--38\,615, 2018.

\bibitem[Zou(2018)]{HetnetSECY2018}
Y.~Zou, ``Intelligent interference exploitation for heterogeneous cellular
  networks against eavesdropping,'' \emph{IEEE Journal on Selected Areas in
  Communications}, vol.~36, no.~7, pp. 1453--1464, 2018.

\bibitem[Liu et~al.(2017{\natexlab{b}})Liu, Qin, Elkashlan, Nallanathan, and
  McCann]{HetnetNOMAYZMAJ2017}
Y.~Liu, Z.~Qin, M.~Elkashlan, A.~Nallanathan, and J.~A. McCann,
  ``Non-orthogonal multiple access in large-scale heterogeneous networks,''
  \emph{IEEE Journal on Selected Areas in Communications}, vol.~35, no.~12, pp.
  2667--2680, 2017.

\bibitem[Liu and Liang(2018)]{HetnetNOMACD2018}
C.-H. Liu and D.-C. Liang, ``Heterogeneous networks with power-domain {NOMA}:
  Coverage, throughput, and power allocation analysis,'' \emph{IEEE
  Transactions on Wireless Communications}, vol.~17, no.~5, pp. 3524--3539,
  2018.

\bibitem[Altay and Koca(2018)]{HetnetNOMACM2018}
C.~Altay and M.~Koca, ``Stochastic geometry analysis of interference
  coordination in {NOMA-based} heterogeneous networks,'' \emph{arXiv preprint
  arXiv:1806.07410}, 2018.

\bibitem[Han et~al.(2018)Han, Gong, Liu, Islam, Li, Bai, and
  Kwak]{HetnetNOMATJXSQZK2018}
T.~Han, J.~Gong, X.~Liu, S.~R. Islam, Q.~Li, Z.~Bai, and K.~S. Kwak, ``On
  downlink {NOMA} in heterogeneous networks with non-uniform small cell
  deployment,'' \emph{IEEE Access}, vol.~6, pp. 31\,099--31\,109, 2018.

\bibitem[Huo et~al.(2019)Huo, Fan, Ma, Cheng, Tian, and
  Chen]{HetnetNOMASECYXLXZD2019}
Y.~Huo, X.~Fan, L.~Ma, X.~Cheng, Z.~Tian, and D.~Chen, ``Secure communications
  in tiered {5G} wireless networks with cooperative jamming,'' \emph{IEEE
  Transactions on Wireless Communications}, vol.~18, no.~6, pp. 3265--3280,
  2019.

\bibitem[Forouzesh et~al.(2019)Forouzesh, Azmi, Mokari, Wong, and
  Pishro-Nik]{HetnetNOMASECMPNKH2019}
M.~Forouzesh, P.~Azmi, N.~Mokari, K.-K. Wong, and H.~Pishro-Nik, ``Robust
  physical layer security for power domain non-orthogonal multiple access-based
  {HetNets} and {HUDNs}: {SIC} avoidance at eavesdroppers,'' \emph{IEEE
  Access}, vol.~7, pp. 107\,879--107\,896, 2019.

\bibitem[Jo et~al.(2012)Jo, Sang, Xia, and Andrews]{HetnetHYPJ2012}
H.-S. Jo, Y.~J. Sang, P.~Xia, and J.~G. Andrews, ``Heterogeneous cellular
  networks with flexible cell association: A comprehensive downlink {SINR}
  analysis,'' \emph{IEEE Transactions on Wireless Communications}, vol.~11,
  no.~10, pp. 3484--3495, 2012.

\bibitem[Andrews et~al.(2011)Andrews, Baccelli, and
  Ganti]{andrews2011tractable}
J.~G. Andrews, F.~Baccelli, and R.~K. Ganti, ``A tractable approach to coverage
  and rate in cellular networks,'' \emph{IEEE Transactions on communications},
  vol.~59, no.~11, pp. 3122--3134, 2011.

\bibitem[Di~Renzo et~al.(2013)Di~Renzo, Guidotti, and Corazza]{HetnetMAG2013}
M.~Di~Renzo, A.~Guidotti, and G.~E. Corazza, ``Average rate of downlink
  heterogeneous cellular networks over generalized fading channels: A
  stochastic geometry approach,'' \emph{IEEE Transactions on Communications},
  vol.~61, no.~7, pp. 3050--3071, 2013.

\bibitem[Anghel and Kaveh(2004)]{equalityAK2004}
P.~A. Anghel and M.~Kaveh, ``Exact symbol error probability of a cooperative
  network in a rayleigh-fading environment,'' \emph{IEEE Transactions on
  Wireless Communications}, vol.~3, no.~5, pp. 1416--1421, 2004.

\bibitem[Jeffrey and Zwillinger(2007)]{BookII2007}
A.~Jeffrey and D.~Zwillinger, ``Table of integrals, series, and products,''
  2007.

\end{thebibliography}
%\begin{thebibliography}{1}

%\end{thebibliography}

\end{document}